\documentclass[a4paper,11pt]{article}
\pdfoutput=1
\usepackage{jheppub}
\usepackage[T1]{fontenc}
\usepackage[latin1]{inputenc}
\usepackage[english]{babel}
\usepackage{amsthm}
\usepackage{empheq}
\usepackage{mathdots}
\DeclareMathOperator{\Tr}{Tr}
\DeclareMathOperator{\sign}{sign}
\DeclareMathOperator{\diag}{diag}
\DeclareMathOperator{\adj}{ad}
\DeclareMathOperator{\rank}{rank}
\usepackage{dsfont}
\usepackage{tikz}
\usetikzlibrary{calc,shapes,arrows,patterns}
\tikzset{>=stealth'}
\newcommand{\HIDDEN}[1]{}
\usepackage[textsize=small,shadow]{todonotes}
\usepackage{ulem}

\title{\boldmath Double Field Theory on Group Manifolds} 
\preprint{LMU-ASC 63/14\\MPP-2014-378}

\author[a,b]{Ralph Blumenhagen,}
\emailAdd{blumenha@mpp.mpg.de}
\author[b]{Falk Hassler,}
\emailAdd{f.hassler@lmu.de}
\author[a,b]{and Dieter L\" ust}
\emailAdd{dieter.luest@lmu.de}

\affiliation[a]{Max-Planck-Institut f\"ur Physik\\
F\"ohringer Ring 6, 80805 M\"unchen, Germany}
\affiliation[b]{Arnold-Sommerfeld-Center f\"ur Theoretische Physik\\
Department f\"ur Physik, Ludwig-Maximilians-Universit\"at M\"unchen\\
Theresienstra\ss e 37, 80333 M\"unchen, Germany}

\abstract{A new version of double field theory (DFT) is derived for the exactly solvable background of an in general left-right asymmetric WZW model in the large level limit. This generalizes the original DFT that was derived via expanding closed string field theory on a torus up to cubic order. The action and gauge transformations are derived for fluctuations around the generalized group manifold background up to cubic order, revealing the appearance of a generalized Lie derivative and a corresponding C-bracket upon invoking a new version of the strong constraint. In all these quantities a background dependent covariant derivative appears reducing to the partial derivative for a toroidal background. This approach sheds some new light on the conceptual status of DFT, its background (in-)dependence and the up-lift of non-geometric Scherk-Schwarz reductions.}

\begin{document}
\maketitle

\section{Introduction}
Dualities are an intriguing property of string theory. They identify the dynamics of a string propagating in two backgrounds which are at a first glance totally different. Nevertheless, for the string these backgrounds are completely indistinguishable. There are two different dualities: S-duality identifies a strongly coupled theory with its weakly coupled counterpart and T-duality which e.g. relates string theories defined on two different tori. Finally, S- and T-duality can be unified into U-duality. Over several years, the study of dualities has revealed some fundamental properties of string theory and has also led to the formulation of M-theory.

Double Field Theory (DFT) is an approach along these lines \cite{Siegel:1993th,Hull:2009mi,Hull:2009zb,Hohm:2010jy,Hohm:2011ex,Aldazabal:2013sca,Hohm:2013bwa}. In order to visualize its significance, consider first supergravity (SUGRA). It describes the target space dynamics of massless closed string excitations and T-duality is only accessible through the Buscher rules \cite{Buscher:1987sk}. However they are non-linear transformations mixing metric and $B$-field, which in general do not correspond to symmetries of the supergravity action. DFT solves this problem by making T-duality a manifest symmetry. It extends the $D$-dimensional target space to a space with $2D$ dimensions called doubled space. In this space a T-duality transformation corresponds to a simple $O(D,D,\mathds{Z})$ rotation. 

DFT was derived from Closed String Field Theory (CSFT) expanding it up to cubic order on a torus\footnote{Much earlier, Siegel derived a doubled theory from 1st-quantized closed string theory \cite{Siegel:1993th}. It is background independent and introduces the strong constraint for the first time.} \cite{Hull:2009mi}. In addition to the $D$ center of mass coordinates $x_i$ of the string, $D$ extra coordinates $\tilde x^i$ were introduced. They are conjugate to the string winding $w_i$, like the coordinates $x_i$ are conjugate to the string momentum $p^i$. The fields on the doubled space are restricted by level matching, a consistency constraint of CSFT. A sufficient condition for closure of the DFT gauge algebra is the strong constraint \cite{Hull:2009zb}. It is more restrictive than level-matching, which is hence also called weak constraint. Equipped with the strong constraint, a background independent version of DFT was derived in \cite{Hohm:2010jy} and shown to be equivalent \cite{Hohm:2010xe} to a theory proposed by Siegel \cite{Siegel:1993th} long before. Its gauge algebra is governed by the C-bracket, which is equivalent to the Courant bracket of Generalized Geometry if the strong constraint holds. Finally, the trivial solution to the strong constraint with vanishing $\tilde x^i$ dependence transforms DFT back into SUGRA. 

Meanwhile, also an extension of DFT was constructed implementing U-duality as a manifest symmetry. It is called Exceptional Field Theory \cite{Hohm:2013vpa,Hohm:2013uia,Hohm:2014fxa,Godazgar:2014nqa} and is constrained by the section condition, a generalization of the strong constraint of DFT.

Thus, it is clear that the strong constraint is a fundamental ingredient of DFT. There are also attempts to soften it, motivated by the fact that it is impossible to obtain all gauged supergravities arising from the embedding tensor formalism (see \cite{Samtleben:2008pe} for a nice introduction) by flux compactifications of SUGRA. Performing a generalized Scherk-Schwarz compactification \cite{Dabholkar:2002sy,Hull:2005hk,Aldazabal:2011nj,Grana:2012rr,Hassler:2014sba} in DFT and substituting the strong constraint by the weaker so-called closure constraint, one is able to reproduce all electrically gauged half-maximal supergravities suggested by the embedding tensor \cite{Dibitetto:2012rk,Geissbuhler:2011mx}. This result suggests that DFT with a weakened constraint is more general than SUGRA. Indeed, the backgrounds related to these gaugings are not accessible from SUGRA and are globally or even locally not well defined. Thus, they are called non-geometric backgrounds. Standard diffeomorphisms and $B$-field gauge transformations are not sufficient to patch them properly. In special cases this problem can be cured by performing a field redefinition \cite{Andriot:2011uh,Andriot:2012wx,Andriot:2012an,Andriot:2013xca,Blumenhagen:2013aia}, but in general it is not possible to describe non-geometric backgrounds in a consistent way in $D$-dimensional target space. Nevertheless, they are totally well defined in the $2D$-dimensional doubled space of DFT with closure constraint. Already before the advent of DFT, the need of a doubled space to treat non-geometric backgrounds was discussed in a series of papers by Hull and Dabholkar \cite{Hull:2004in,Dabholkar:2005ve,Hull:2006va}.

Motivated by these findings, the flux formulation of DFT was developed \cite{Geissbuhler:2013uka,Aldazabal:2013sca}. Up to total derivatives it is equivalent to the original formulation after applying the strong constraint. In general, the flux formulation substitutes the strong constraint by the closure constraint so that additional terms are allowed in the action. Furthermore, all dynamical fields are encoded in the $O(D,D)$ covariant fluxes $\mathcal{F}_{ABC}$. These are equivalent to the embedding tensor mentioned above. 

The picture outlined by these developments shows the power of dualities: Starting from a reformulation to make T-duality manifest, one is allowed to go beyond well known geometric string backgrounds. However often for this general approach, the uplift to string theory and conformal field theory is not clear. There are examples related to asymmetric orbifolds \cite{Dabholkar:2002sy,Condeescu:2012sp,Condeescu:2013yma,Hassler:2014sba} which provide some evidence that at least for these cases uplifts to string theory exist.

In this paper we intend to provide a new perspective upon the traditional version of DFT, in particular on the issues related to the strong constraint, background (in-)dependence and uplifts of non-geometric configurations. For that purpose, we are going back to the root of DFT and 
evaluate the CSFT action up to cubic order for a non-toroidal solution to the string equations of motion. Indeed, instead of considering a flat torus as the background, like in the work of Hull and Zwiebach \cite{Hull:2009mi}, we use a string propagating on a compact group manifold with fluxes. Due to their isometries, these manifolds have the same local properties at each point. Generically, these isometries are non-abelian, but they include also the torus with abelian isometries. Group manifold are also well suited to study various properties of doubled geometries \cite{Dabholkar:2005ve,Hull:2009sg}.

On the world-sheet, the exactly solvable background is described by a Wess-Zumino-Witten model (WZW) \cite{Witten:1983ar} in the large radius/level limit $(k\gg 1)$. Employing the occurring current algebras, we derive a cubic action and the corresponding gauge transformations from CSFT. Just like in DFT, we find that one also has to impose a weak/strong constraint, which however takes a different form. Instead of partial derivatives, it exhibits additional terms which can be adsorbed into a connection forming a covariant derivative. The same pattern also appears for the generalized Lie derivative and the C-bracket. Therefore, the gauge algebra we derive resembles the one proposed by Cederwall \cite{Cederwall:2014kxa} with the difference that the connection encountered in our approach turned out not to be torsion-free.

Due to the split into left- and right movers, the gauge algebra closes even for asymmetric backgrounds, i.e. for backgrounds not solving the traditional strong constraint of DFT. Thus, our set-up is general enough to describe fluctuation around backgrounds that violate the strong constraint and, in this respect, goes beyond the framework of traditional DFT. These asymmetric WZW backgrounds, at least in the large radius/level limit, are candidates for the uplift of non-geometric configurations beyond the well studied locally flat asymmetric (toroidal) orbifold examples. All these findings suggest that the theory we derive in this paper is a generalization of traditional DFT, though containing it for a toroidal background. In order to distinguish them, we call it ${\rm DFT}_{\rm WZW}$.

This paper is organized as follows: In section~\ref{sec:worldsheet}, we review the relevant features of the WZW model and its current algebra. Furthermore, we give a representation for two- and three-point correlators involving these currents in terms of scalar functions on a group manifold in the limit of large level $k$. Section~\ref{sec:CSFTaction&gauge} presents the derivation of the action and its gauge transformations to cubic order in CSFT. In section~\ref{sec:cbracket} we discuss the generalized Lie derivative, the gauge algebra and the constraint necessary for its closure. Finally in section~\ref{sec:thoughts}, we start an investigation of the relation between the theory constructed in this paper so far and the flux formulation of toroidal DFT. There, we also propose the string theory uplift of non-geometric flux backgrounds arising from a generalized Scherk-Schwarz
compactification of traditional DFT. 

\section{World-sheet theory for strings on a group manifold}\label{sec:worldsheet}
In this section, we briefly review the WZW model and its current algebra providing the notation for the rest of the paper. For a more detailed review of WZW models, we refer to e.g. \cite{Walton:1999xc} or the appendix of \cite{Schulz:2011ye}. Additionally, we show how the various representations of a semisimple Lie algebra can be expressed in terms of scalar functions on the group manifold. Afterwards, we use this result to express two- and three-point correlators and show that they fulfill the Knizhnik-Zamolodchikov equation \cite{Knizhnik:1984nr}. Finally, we provide the two- and three-point off-shell amplitude for Ka\v{c}-Moody primary fields. 

\subsection{Wess-Zumino-Witten model and Ka\v{c}-Moody current algebra}\label{sec:wzwmodel}
A string propagating on a group manifold of a semisimple Lie group $G$ is described by the non-linear sigma model
\begin{equation}\label{eqn:WZWaction}
  S = \frac{1}{4\pi\alpha'} \int_{\partial M} \mathcal{K}(\omega_\gamma, \star \omega_\gamma) + S_\mathrm{WZ}
\end{equation}
on the world-sheet two-sphere $S^2=\partial M$. Note that its prefactor does not match the common choice $-k/(8\pi)$, but it is very convenient for comparing \eqref{eqn:WZWaction} with a non-linear sigma model given in terms of a metric and an asymmetric two-form field. We will compensate for this uncommon choice in the definition of the Killing metric \eqref{eqn:killingform}. The action given here is exactly the same as the one presented in \cite{Walton:1999xc}.

Let us explain the notation used in \eqref{eqn:WZWaction} in more detail. As usual, $\star$ denotes the Hodge dual and $\omega_\gamma$ is the left-invariant Maurer-Cartan form\footnote{We could also use the right-invariant Maurer-Cartan form and would obtain the same results. But in the literature it is common to use the left-invariant one.}. The function $\gamma(\sigma)$, which appears as subscript of $\omega_\gamma$, maps each point of $S^2$ to an element of the group $G$. In this way the string world sheet is embedded into the target space. In order to fix a certain group element $\gamma \in G$, one needs $D$ different parameters $x^i$ where $i$ runs from one to $D$. Infinitesimal changes of them at a fixed $\gamma$ create the tangent space $T_\gamma G$ of the group manifold. At the identity, $T_e G$ is identified with the Lie algebra $\mathfrak{g}$ associated to $G$. The tangent space at an arbitrary group element $T_\gamma$ is mapped to $\mathfrak{g}$ by the left- or right-invariant Maurer-Cartan form
\begin{equation}\label{eqn:maurer-cartan}
  \omega_\gamma = \gamma^{-1} d \gamma = \gamma^{-1} \partial_i \gamma \,d x^i \quad \text{or} \quad
  \bar\omega_\gamma = d \gamma \gamma^{-1} = \partial_i \gamma \gamma^{-1} \, d x^i \quad  \text{with} \quad
  \partial_i = \frac{\partial}{\partial x^i}\,.
\end{equation}
They arise if $\gamma$ is assumed to act as a left or right translation of
$G$. Both of them take values in the Lie algebra $\mathfrak{g}$. Two elements
of this algebra are contracted to a scalar by the symmetric, bilinear Killing
form\footnote{We use the common convention that the length square of the longest root in the root system of $\mathfrak{g}$ is normalized to 2.}
\begin{equation}\label{eqn:killingform}
  \mathcal{K}(x, y) = - \frac{\alpha' k}{2} \frac{\Tr (\adj_x \adj_y )}{2 h^\vee}\,,
    \quad \text{with} \quad
  x,\,y\in \mathfrak{g}
\end{equation}
where $\adj_x$ is the adjoint representation of $x$ and $h^\vee$ denotes the
dual Coxeter number of $\mathfrak{g}$. The generalization of this equation to
$n$-forms is straightforward: One has to insert a wedge product $\wedge$ between $\adj_x$ and $\adj_y$. With these definitions at hand, one is able to expand \eqref{eqn:WZWaction} as
\begin{equation}\label{eqn:WZWactionmetric}
  S = \frac{1}{4\pi\alpha'} \int_{\partial M} g_{ij} \, d x^i \wedge \star d x^j + S_\mathrm{WZ} 
    \quad \text{with} \quad 
  g_{ij} = \mathcal{K}(\gamma^{-1}\partial_i \gamma, \gamma^{-1}\partial_j \gamma)
\end{equation}
where $g_{ij}$ is the target space metric of the group manifold. The parameters $x^i$ parameterizing the elements of the group $G$ are equivalent to coordinates on the manifold. They are related to the word-sheet coordinates $\sigma^\alpha$ by the mapping $x^i(\sigma^a)$ giving rise to $d x^i = \partial_\alpha x^i d\sigma^\alpha$.

Since the metric part \eqref{eqn:WZWactionmetric} of the action $S$ alone spoils local conformal symmetry, one has to add the topological Wess-Zumino term
\begin{gather}\label{eqn:SWZ}
  S_\mathrm{WZ} = \frac{1}{12 \pi \alpha'} \int_M  \mathcal{K}\left(\omega_\gamma, 
    [ \omega_\gamma, \omega_\gamma ]\right) = \frac{1}{2 \pi \alpha'} \int_M H \\
\intertext{with the 3-form flux}\label{eqn:WZWHflux}
  H = \frac{1}{3!} H_{ijk} \, d x^i \wedge d x^j \wedge d x^k
    \quad \text{and} \quad
  H_{ijk} = \mathcal{K}\left(\gamma^{-1} \partial_i \gamma, [\gamma^{-1} \partial_j \gamma, \gamma^{-1} \partial_k \gamma]\right)\,.
\end{gather}
Here, the $H$-flux is the field strength associated to the massless, antisymmetric Kalb-Ramond field $B_{ij}$. Both are linked via the relation\footnote{Within this paper we use the notation
\begin{equation*}
  T_{[a_1 \dots a_n]} = \frac{1}{n!}  \sum\limits_{\sigma\in P} \sign(\sigma) T_{\sigma_1 \dots \sigma_n}
  \quad \text{and} \quad 
  T_{(a_1 \dots a_n)} = \frac{1}{n!}  \sum\limits_{\sigma\in P} 
    T_{\sigma_1 \dots \sigma_n}\,,
\end{equation*}
to denote the (anti)symmetrization of rank $n$ tensors. $P$ is the set of all permutations of the indices $a_1,\dots,a_n$.}
\begin{equation}
  H = d B \quad \text{with} \quad 
  B = \frac{1}{2!} B_{ij} d x^i \wedge d x^j 
  \quad \text{and} \quad
  H_{ijk} = 3 \partial_{[i} B_{jk]}\,.
\end{equation}
Of course, a physically meaningful sigma model only depends on the world sheet $\partial M$ and not on its extension to the three-dimensional space $M$. Thus, physics has to be independent of the specific choice for $M$. For $G$ being a compact semisimple Lie groups with non-trivial homology $\pi_3(G)=\mathds{Z}$, this is only the case if $S_\mathrm{WZ}$ is an integer multiple of $2\pi$ \cite{Witten:1983tw}. Thereby, the $H$-flux of a compact background is quantized.

The variation of the action with respect to the $G$-valued field $\gamma$ gives rise to the equation of motion
\begin{equation}\label{eqn:eomwzwmodel}
  \partial_\alpha (\gamma^{-1} \partial^\alpha \gamma) + \frac{1}{2} \epsilon_{\alpha\beta} \partial^\alpha (\gamma^{-1} \partial^\beta \gamma) = 0\,.
\end{equation}
It is interesting to note that the second term in this equation origins from the Wess-Zumino term in the action. By fixing the word sheet metric to $h^{z\bar z} = 2$, $h^{zz}=h^{\bar z\bar z}=0$ and writing out the components of the totally antisymmetric tensor $\epsilon_{\alpha\beta}$ with $\epsilon_{z \bar z}=1$, one obtains
\begin{equation}
  \partial (\gamma^{-1} \bar \partial \gamma) = 0 \,.
\end{equation}
One can directly read off the anti-chiral Noether current
\begin{equation}\label{eqn:antichiralcurrent}
  \bar j(\bar z) = - \frac{2}{\alpha'} \gamma^{-1} \bar \partial \gamma
\end{equation}
from the equation of motion. Note that, without the second term in \eqref{eqn:eomwzwmodel}, we would not obtain an anti-holomorphic current. To obtain the chiral current, we apply complex conjugation to \eqref{eqn:antichiralcurrent} and substitute $\gamma$ by $\gamma^{-1}$ afterwards. By this procedure we get
\begin{equation}\label{eqn:chiralcurrent}
  j(z) = \frac{2}{\alpha'} \partial \gamma \gamma^{-1}  \,.
\end{equation}
To motivate the normalization of these currents, consider the infinitesimal transformations
\begin{equation}\label{eqn:deltaxigamma}
  \delta_\xi \gamma(w,\bar w) = \xi(w) \gamma(w,\bar w) 
    \quad \text{and} \quad
  \delta_{\bar \xi} \gamma(w,\bar w) = - \gamma(w,\bar w) \bar \xi(\bar w) \,.
\end{equation}
of the field $\gamma$. Here, $\xi(w)$ and $\bar \xi(\bar w)$ are the Lie algebra valued parameters of the transformations. It is sufficient to discuss the chiral part $\xi(z)$ only. Applying \eqref{eqn:deltaxigamma} to the action $S$, we obtain 
\begin{equation}
  \delta_\xi S = - \frac{1}{2\pi i} \oint_0 dz\, \mathcal{K}(\xi(z), j(z))
\end{equation}
where $\oint_w dz$ denotes a closed contour integral around the point $w$. Here, we have chosen the normalization factor of $j_a$ in \eqref{eqn:chiralcurrent} to obtain precisely the factor $1/(2\pi i)$ in this expression. With $\delta S$ one can  compute small changes
\begin{equation}\label{eqn:deltaxi<x>}
  \delta_\xi \langle X \rangle = \langle \delta_\xi S X \rangle = \frac{1}{2\pi i} \oint_0 dz \langle \mathcal{K}(\xi(z), j(z)) X \rangle
\end{equation}
of an arbitrary expectation value
\begin{equation}
  \langle X \rangle = \frac{\int [d \gamma]\, X e^{-S[\gamma]}}{\int [d \gamma]\, e^{-S[\gamma]}}
\end{equation}
in the Euclidean path integral. 

As a brief interlude, let us discuss the $D = \dim\mathfrak{g}$ generators $t_a$ of the Lie algebra $\mathfrak{g}$. They form a basis of the adjoint representation. We define the symmetric tensor
\begin{equation}\label{eqn:defetaab}
  \eta_{ab} = \mathcal{K}(t_a, t_b) = - \frac{\alpha' k}{2} \frac{\Tr(t_a t_b)}{2 x_\lambda} = - \frac{1}{2 h^\vee}
    f_{ad}{}^c f_{bc}{}^d \,.
\end{equation}
In the last step we have expressed the generators in terms of the structure coefficients of the Lie algebra appearing in the commutation relation\footnote{There are different conventions. Some use an additional $i$ in front of the structure coefficients. We stick to the convention in \cite{Schulz:2011ye} without $i$.}
\begin{equation}\label{eqn:liealgebra}
  [t_a, t_b] = \sqrt{\frac{2}{\alpha' k}} f_{ab}{}^c\, t_c = F_{ab}{}^c\, t_c
    \quad \text{with} \quad
  F_{ab}{}^c := \sqrt{\frac{2}{\alpha' k}}\, f_{ab}{}^c  \,. 
\end{equation}
For later convenience, we defined the rescaled structure coefficients $F_{ab}{}^c$. Note that it is always possible to choose the generators $t_a$ of a semisimple Lie algebra $\mathfrak{g}$ in a way that $\eta_{ab}$ is a diagonal matrix with entries $\pm 1$. Thus, $\eta_{ab}$ is completely specified by its signature. A compact Lie group $G$  has a Lie algebra with a negative definite Killing form, i.e. the signature of $\eta_{ab}$ is $(-, \dots,-)$. In combination with its inverse $\eta^{ab}$, $\eta_{ab}$ is used to raise and lower flat indices $a,b,\dots$. 

Coming back, the chiral current \eqref{eqn:chiralcurrent} can be written in terms of the generators $t_a$ as
\begin{equation}
  j(z) = t^a j_a(z) \quad \text{with} \quad j_a(z) = \mathcal{K}(t_a, j(z)) \,.
\end{equation}
In this form, the infinitesimal transformation $\delta_\xi$ of the chiral current reads
\begin{equation}\label{eqn:deltaxij}
  \delta_\xi  j_b(w) = F_{ab}{}^c\, j_c(w)\, \xi^a(w)  + \frac{2}{\alpha'} \eta_{ab} \partial \xi^a(w) 
    \quad \text{with} \quad
  \xi_a(w) = \mathcal{K}(t_a, \xi(w))\,.
\end{equation}
Plugging this into \eqref{eqn:deltaxi<x>} one obtains
\begin{equation}
  \delta_\xi \langle j_b(w) \rangle = \frac{1}{2\pi i} \oint d z \langle j_a(z) j_b(w) \rangle \xi^a(z) 
    = F_{ab}{}^c\, \langle j_c(w) \rangle \xi^a(w) + \frac{2}{\alpha'} \eta_{ab} \partial \xi^a(w)
\end{equation}
allowing to read off the OPE
\begin{empheq}[box=\fbox]{equation}\label{eqn:opejj}
  j_a(z) j_b(w) = \frac{F_{ab}{}^c\, j_c(w)}{z-w} - \frac{2}{\alpha'} \frac{\eta_{ab}}{(z-w)^2} + \dots
\end{empheq}
of the chiral currents. The analogous algebra holds for the anti-chiral current $\bar j(\bar z)$. Normally one would expect the level $k$ in front of the flat metric $\eta_{ab}$ instead of $-\alpha'/2$. Here, $k$ is hidden in the rescaled structure coefficients $F_{ab}{}^c$. For this reason, the OPE \eqref{eqn:opejj} corresponds to the usual form of the Ka\v{c}-Moody algebra at level $k$. Applying the same procedure to the transformation in \eqref{eqn:deltaxigamma}, we get the OPE
\begin{empheq}[box=\fbox]{equation}\label{eqn:opejg}
  j_a(z) \gamma(w,\bar w) = \frac{t_a \gamma(w,\bar w)}{z-w} + \cdots
\end{empheq}
defining a Ka\v{c}-Moody primary. Introducing the mode expansion
\begin{equation}\label{eqn:modeexpja}
  j_a(z) = \sum\limits_n j_{a,n} \, z^{-n-1}
\end{equation}
the OPE \eqref{eqn:opejj} is equivalent to the Ka\v{c}-Moody algebra  
\begin{equation}\label{eqn:algebrajj} 
  [j_{a,m}, j_{b,n}] = F_{ab}{}^c\, j_{c,m+n} - \frac{2}{\alpha'}\, m\, \eta_{ab}\, \delta_{m+n} \,.
\end{equation}

\subsection{A geometric representation for semisimple Lie algebras}\label{sec:representation}
In the following we will show that there exist highest weight representations of a semi-simple Lie algebra in terms of scalar functions defined on the group manifold. For that purpose, let us first change from the abstract notation with Maurer-Cartan forms to a more explicit one by introducing vielbeins. Expressing $\omega_\gamma$ in \eqref{eqn:maurer-cartan} in terms of the generators $t_a$, we obtain
\begin{equation}\label{eqn:vielbein}
  \omega_\gamma = t_a \, e^a{}_i \, d x^i \quad \text{with the vielbein} \quad
    e^a{}_i = \mathcal{K}(t^a, \gamma^{-1} \partial_i \gamma)\,.
\end{equation}
It carries two different kinds of indices: flat ones are labeled by $a,b,c,\cdots$ and curved ones by $i,j,k,\cdots$. Flat indices are raised and lowered with the metric $\eta_{ab}$, whereas for curved indices we use the target space metric $g_{ij}$ in \eqref{eqn:WZWactionmetric}, which in terms of the vielbein reads
\begin{equation}
  g_{ij} = \eta_{ab}\, e^a{}_i\, e^b{}_j \,.
\end{equation}
Moreover, $e_a{}^i$ denotes the inverse transposed of $e^a{}_i$ and the $H$-flux defined in \eqref{eqn:WZWHflux} can be written as
\begin{equation}\label{eqn:WZWHfluxexplicit}
  H_{ijk} = e^a{}_i\, e^b{}_j\, e^c{}_k\, F_{abc}\, .
\end{equation}
Introducing the flat derivative
\begin{equation}\label{eqn:flatderivative}
  D_a = e_a{}^i \partial_i
\end{equation}
the commutator of two of them satisfies
\begin{equation}
  [D_a, D_b] = F_{ab}{}^c D_c \, ,
\end{equation}
with
\begin{equation}\label{eqn:fabcfromvielbein}
  F_{ab}{}^c = 2 e_{[a}{}^i \partial_i e_{b]}{}^j e^c{}_j = 2 D_{[a} e_{b]}{}^i e^c{}_i\,.
\end{equation}
Thus, we found a representation of the generators $t_a$ in terms of the differential operators $D_a$ acting on functions defined on a patch of the group manifold. We will see that these functions include all highest weight  representations of the Lie algebra.

Flat derivatives are mainly used under volume integrals with the volume element $d^D x \sqrt{|g|}$ where $g$ denotes the determinate of the target space metric $g_{ij}$. In this case, one finds
\begin{equation}\label{eqn:inttotalderiv}
  \int d^D x \, \sqrt{|g|} D_a v= \int d^D x \, \partial_i(\sqrt{|g|} e_a{}^i v)\,,
\end{equation}
where $v$ is an arbitrary scalar function depending on the target space coordinates $x^i$. Thus, the right hand side reduces to a boundary term, which we always assume to vanish. Then one can perform integration by parts
\begin{equation}
  \int d^D x \, \sqrt{|g|} (D_a v) w = - \int d^D x \, \sqrt{|g|} v (D_a w)\,.
\end{equation}
Note that \eqref{eqn:inttotalderiv} is not restricted to semisimple Lie algebras, but is much more general and always holds if
\begin{equation}\label{eqn:unimodular}
  F_{ab}{}^b = 0 \quad \text{or equivalently} \quad \Tr \adj_x = 0 \quad \forall x \in \mathfrak{g}
\end{equation}
is fulfilled. Lie algebras with this property are called unimodular. 

The well known procedure of building highest weight representations also carries over to the flat derivatives discussed above. Take e.g. the group $SU(2)$ parameterized by Hopf coordinates $x^i = (\eta^1, \eta^2, \eta^3)$ with $0\le\eta^1<\pi/2$ and $0\le\eta^{2,3}<2\pi$. A detail derivation of the vielbeins for this group is presented in appendix~\ref{app:su(2)}. Here we are only interested in the flat derivatives
\begin{align}
  \tilde D_3 &= - \sqrt{\frac{\alpha' k}{2}} D_3 = - \frac{i}{\sqrt{2}}\bigl( \partial_2 + \partial_3 \bigr) \quad \text{and} \\
  \tilde D_\pm &= - \sqrt{\frac{\alpha' k}{2}} ( \pm i D_1 - D_2 ) \nonumber\\ 
    &= - \frac{i e^{\pm i(\eta^2+\eta^3)}}{\sqrt{2} \sin(2 \eta^1)} \left[ \pm i \sin(2 \eta^1) \,\partial_1 + 2 \sin^2(\eta^1)\, \partial_2 - 2 \cos^2(\eta^1)\, \partial_3 \right]\,.
\end{align}
We look for eigenfunctions of $\tilde D_3$ which are annihilated by $\tilde D_+$. A short calculation shows that this is the case for
\begin{equation}
  y_\lambda(x^i) = C_\lambda ( \sin \eta^1 )^{\sqrt{2} \lambda} e^{i \sqrt{2} \lambda \eta^3}
\end{equation}
where $C_\lambda$ denote normalization constant constants fixed by the requirement
\begin{equation}\label{eqn:normalizationy}
  \int d^D x\, \sqrt{|g|} y_\lambda^* y_\lambda^{} = |C_\lambda|^2 4 \pi^2 (\alpha' k)^{3/2} \int\limits_0^{\pi/2} d \eta^1 \, 
  \cos(\eta^1) \sin(\eta^1)^{1 + 2 \sqrt{2} \lambda} = |C_\lambda|^2
  \frac{2 \pi^2 (\alpha' k)^{3/2}}{\sqrt{2} \lambda + 1} = 1\, ,
\end{equation}
which is only possible if $\sqrt{2} \lambda + 1 > 0$. Furthermore, we know from $\mathfrak{su}(2)$ representation theory that $\lambda$ is an element of the 1-dimensional weight lattice $\Lambda = \mathds{Z}/\sqrt{2}$.Therefore, $\lambda$ has to be an element of $\mathds{N}_0/\sqrt{2}$ in order to allow the normalization \eqref{eqn:normalizationy}. Starting from these highest weight states, one can construct the full $\mathfrak{su}(2)$ representation by acting with $\tilde D_-$ on $y_\lambda$. We denote the resulting functions according to their $\tilde D_3$ eigenvalues as
\begin{equation}
  y_{\lambda q} = C_{\lambda q} (\tilde D_-)^{(\lambda - q)/\sqrt{2}} y_\lambda \quad \text{with} \quad
  \tilde D_3 y_{\lambda q} = q\, y_{\lambda q}\quad \text{and} \quad
  q=-\lambda, -\lambda + \sqrt{2}, \dots, \lambda \,.
\end{equation}
Some of these functions are listed in appendix~\ref{app:su(2)}. According to the integral
\begin{equation}
  \int d^D x \sqrt{|g|}\, y_{\lambda_1 q_1}^*\, y_{\lambda_2 q_2}^{} = \delta_{\lambda_1 \lambda_2} \delta_{q_1 q_2}\,,
\end{equation}
which fixes the normalization constants $C_{\lambda q}$, they form an orthonormal basis of the Hilbert space of square-integrable functions on the 3-sphere $L_2(S^3)$. It is straightforward to generalized this procedure for other compact semisimple Lie algebras. In this case $\lambda$ and $q$ are not just scalars, but vectors of dimension $r = \rank \mathfrak{g}$.

For non-compact Lie algebras, the structure becomes more involved: First, one has to consider lowest weight states in addition to the highest weight states discussed so far. These are states annihilated by all negative simple roots. A representation is build by acting with all negative simple roots on highest weight states $v_\lambda$ and with all positive simple roots on lowest weight states $v_{-\lambda}$. In contrast to a compact Lie algebra, this process does not terminate. Thus, there is an infinite tower of states for each highest and lowest weight. A simple example for a non-compact Lie algebra is $\mathfrak{sl}(2)$. Its representations are discussed in the context of the $SL(2)$ WZW model in \cite{Maldacena:2000hw}. 

\subsection{Two- and three-point correlation functions of Ka\v{c}-Moody primaries}\label{sec_twothreepointkm}
In order to perform the CSFT calculation in the next section, we need to know the correlation functions $\langle \gamma_1(w_1) \dots \gamma_n(w_n)\rangle$ of Ka\v{c}-Moody primary fields. We have already defined their OPE in \eqref{eqn:opejg}. These have to fulfill the Knizhnik-Zamolodchikov equation \cite{Knizhnik:1984nr}
\begin{equation}\label{eqn:KZeq}
  \left(\partial_{w_i} + \frac{2}{\alpha'} \frac{k}{k + h^\vee} \sum\limits_{i\ne j} \frac{\eta^{ab}\, t_a^{(i)} \otimes t_b^{(j)}}{w_i - w_j} \right) \langle \gamma_1(w_1) \dots \gamma_n(w_n)\rangle = 0
\end{equation}
where the notation $t_a^{(i)}$ indicates that the generator $t_a$ acts on the $i$th field $\gamma_i(w_i)$. The chiral energy momentum tensor is given by the Sugawara construction as
\begin{equation}\label{eqn:T(z)}
  T(z) = -\frac{\alpha'}{2} \frac{k}{2(k + h^\vee)} :\eta^{ab} j_a(z) j_b(z): \,.
\end{equation}
Again, the uncommon factors in the Knizhnik-Zamolodchikov equation and the energy momentum tensor are due to the normalization we performed in section~\ref{sec:worldsheet}. With the OPE of the chiral currents $j_a(z)$ in \eqref{eqn:opejj}, it is straightforward to calculate
\begin{align}
  T(z) j_a(w) &= \frac{j_a(w)}{(z-w)^2} + \frac{\partial_w j_a(w)}{z-w} + \dots 
    \quad \text{and} \quad \\
    T(z) T(w) &= \frac{c}{2 (z-w)^4} + \frac{2\, T(w)}{(z-w)^2} + \frac{\partial_w T(w)}{z-w} + \dots
\end{align}
with the central charge
\begin{equation}\label{eqn:centralcharge}
  c  = \frac{k D}{k + h^\vee}
    \quad \text{and} \quad D = \dim \mathfrak{g}\,.
\end{equation}
From there, one can compute the OPE
\begin{equation}\label{eqn:opeTg}
  T(z) \gamma(w) = \frac{h}{(z-w)^2} \gamma(w) + \frac{\partial_w \gamma(w)}{z-w}  + \dots 
    \quad \text{with} \quad
  h = -\frac{\alpha' k}{4(k + h^\vee)}t_a t^a\,.
\end{equation}
For $\gamma(w)$ to be a Ka\v{c}-Moody and a Virasoro primary, it needs
to be an eigenstate of the Lie algebra's quadratic Casimir operator $\eta^{ab} t_a t_b$.

The CSFT calculation in this paper will be performed only up to quartic order so that we need to know the two-point and three-point correlation functions. Recall that for Virasoro primaries, these are completely determined up to some structure constants. We introduce a Fourier-type expansion of the Ka\v{c}-Moody primary
\begin{equation}
\label{gammaexpand}
  \gamma(w) = \sum\limits_{\lambda, q} c_{\lambda q}\, \phi_{\lambda q}(w, x^i)
\end{equation}
in terms of the Virasoro primaries $\phi_{\lambda q}(w, x^i)$ with constant coefficients $c_{\lambda q}$. Due to the linearity of the correlation functions, it is sufficient to know the correlations functions of $\phi_{\lambda q}$. As mentioned above, these are fixed by conformal symmetry as
\begin{align}\label{eqn:conformal2point}
  \langle \phi_{\lambda_1 q_1}(w_1) \phi_{\lambda_2 q_2}(w_2) \rangle &=
    \frac{d_{\lambda_1 q_1 \, \lambda_ 2 q_2}
    \delta_{h_{\lambda_1} h_{\lambda_2}}}{w_{12}^{2 h_{\lambda_1}}} \quad
    \text{with} \quad w_{12} = w_1 - w_2\, ,\\
  \label{eqn:conformal3point}
  \langle \phi_{\lambda_1 q_1}(w_1) \phi_{\lambda_2 q_2}(w_2) \phi_{\lambda_3 q_3}(w_3) \rangle &=
    \frac{C_{\lambda_1 q_1\,\lambda_2 q_2\,\lambda_3 q_3}}{w_{12}^{h_{\lambda_1}+h_{\lambda_2}-h_{\lambda_3}}
    w_{23}^{h_{\lambda_2}+h_{\lambda_3}-h_{\lambda_1}} w_{13}^{h_{\lambda_1}+h_{\lambda_3}-h_{\lambda_2}}}\,.
\end{align}
In these equations, $h_\lambda$ denotes the conformal weight of $\phi_{\lambda q}$ as written in \eqref{eqn:opeTg}. Note that it is independent of $q$. 

Finally, we apply the Knizhnik-Zamolodchikov equation \eqref{eqn:KZeq} to fix the constants $d_{\lambda_1 q_1 \,\lambda_2 q_2}$ and $C_{\lambda_1 q_1\, \lambda_2 q_2\, \lambda_3 q_3}$ in \eqref{eqn:conformal2point} and
\eqref{eqn:conformal3point}. To do so, we realize that the functions $y_{\lambda q}(x^i)$ we introduced in the last section are eigenstates of
$L_0$, too. Therefore, a natural candidate for the two-point structure constant is
\begin{equation}
  d_{\lambda_1 q_1 \, \lambda_2 q_2} = \int d^D x \sqrt{|g|}\, y_{\lambda_1 q_1}^*\, y_{\lambda_2 q_2}^{} = \delta_{\lambda_1 \lambda_2} \delta_{q_1 q_2}\,.
\end{equation}
We now show that this is compatible with the Knizhnik-Zamolodchikov equation. It automatically implies the delta function $\delta_{h_{\lambda_1} h_{\lambda_2}}$ in \eqref{eqn:conformal2point} by its $\delta_{\lambda_1 \lambda_2}$ part. Plugging the correlation function into \eqref{eqn:KZeq} gives rise to
\begin{equation}
  h_{\lambda_1} d_{\lambda_1 q_1\, \lambda_2 q_2} - \frac{\alpha'}{2}
  \frac{k}{2(k + h^\vee)} \int d^D x\, \sqrt{|g|} \,\tilde D_a y_{\lambda_1
    q_1}^*\,  \tilde D^a y_{\lambda_2 q_2}^{} = 0 \,.
\end{equation}
where we used that the differential operators $\tilde D_a$ give a representation of the Lie algebra generators $t_a$. Now, we perform integration by parts, pull the factor in front of the integrand and obtain
\begin{equation}
  h_{\lambda_1} d_{\lambda_1 q_1\, \lambda_2 q_2} - \int d^D x \sqrt{|g|}\; L_0\, y_{\lambda_1 q_1}^*\, y_{\lambda_2 q_2}^{} = 0\,.
\end{equation}
Recalling the eigenvalue equation $L_0\, y_{\lambda q} = h_\lambda\, y_{\lambda
  q}$, one immediately sees that the Knizhnik-Zamolodchikov equation is indeed fulfilled. A similar calculation proofs that in order to fulfill
\eqref{eqn:KZeq} for the three-point correlation function \eqref{eqn:conformal3point}, we have to set
\begin{equation}\label{eqn:Cconformal3point}
  C_{\lambda_1 q_1\,\lambda_2 q_2\,\lambda_3 q_3} = \int d^D x\, \sqrt{|g|}\, y_{\lambda_1 q_1}^*\, y_{\lambda_2 q_2}^{ }\, y_{\lambda_3 q_3}^{ }\,.
\end{equation}

Let us discuss how the usual toroidal case fits into this scheme. A torus corresponds to an abelian group manifold with $F_{ab}{}^c = 0$ and a coordinate independent vielbein $e_a{}^i$. Applied to the torus metric $g_{ij}=\delta_{ij}$, it gives rise to the flat metric $\eta_{ab}= e_a{}^i g_{ij} e_b{}^j$. Plugging these quantities in \eqref{eqn:algebrajj} and introducing the abelian currents 
\begin{equation}\label{eqn:alpha}
  \alpha_{i,m} = - i \sqrt{\frac{\alpha'}{2}} e^a{}_i\, j_{a,m}\,,
\end{equation}
we obtain the same current algebra
\begin{equation}
  [\alpha_{i, m}, \alpha_{j,n}] = m\, g_{ij}\, \delta_{m+n}
\end{equation}
as used for the derivation of DFT on a torus in \cite{Hull:2009mi}. To reproduce the zero mode $\alpha_{i,0}$, we perform the substitution $j_{a,0} \rightarrow D_a$ giving rise to
\begin{equation}
  \alpha_{i,0} = - i \sqrt{\frac{\alpha'}{2}} D_i \,.
\end{equation}
Finally, the Virasoro zero mode read
\begin{equation}
  L_0 = -\frac{\alpha'}{4} \eta^{ab} \sum_n : j_{a,n}\, j_{b,-n}: = N +
  \frac{1}{2} g^{ij}\, D_i \, D_j
    \quad \text{with} \quad
  N = \sum\limits_{n>0} g^{ij}\, \alpha_{i,n} \alpha_{j,-n}\,.
\end{equation}

Note that the operator $D_a D^a$ is the Laplace operator on the group manifold. As we have seen above, the functions $y_{\lambda q}$ are its eigenfunctions. Consider now flat space where we find
\begin{equation}
  y_k(x^i) = \frac{1}{\sqrt{2 \pi}} e^{i k_i x^i}
\end{equation}
as eigenfunctions of the Laplace operator. The corresponding expansion  \eqref{gammaexpand} is nothing else than a Fourier expansion. According to \eqref{eqn:Cconformal3point}, the constant in the three-point correlation function reads
\begin{equation}
  C_{k_1 \, k_2 \, k_3} = \delta_{- k_1 + k_2 + k_3}\,.
\end{equation}
Physically, this reflects  momentum conservation in a scattering process with two incoming particles (momentum $k_2$ and $k_3$) and one outgoing particle (momentum $k_1$). Switching to the $SU(2)$ example discussed in appendix~\ref{app:su(2)}, one obtains \cite{WenAvery85}
\begin{equation}
  C_{\lambda_1 q_1\,\lambda_2 q_2\,\lambda_3 q_3} = \langle j_1 q_1 | j_2 q_2 \, j_3 q_3 \rangle
\end{equation}
with $\langle j_1 q_1 | j_2 q_2 \, j_3 q_3 \rangle$ denoting the Clebsch-Gordan coefficients. In contrast to flat space, the corresponding scattering process is not ruled by momentum conservation but by angular momentum conservation.

\subsection{Doubled space and fundamental CSFT off-shell amplitudes}\label{sec:fundamentalamp}
In the previous subsection we considered only the chiral primary $\phi_{\lambda q}(w)$. Now, we take also their anti-chiral counterparts $\bar \phi(\bar w)_{\bar \lambda \bar q}$ into account. In order to keep the notation as simple as possible, we introduce the following abbreviations:
\begin{equation}
  R = (\lambda q\,, \bar\lambda \bar q) \quad \text{and} \quad \phi_R(w, \bar w) = \phi_{\lambda q}(w) \bar\phi_{\bar \lambda \bar q}(\bar w)\,.
\end{equation} 
For the WZW model in section~\ref{sec:wzwmodel}, the anti-chiral current $\bar j_a(\bar z)$ is governed by the same Ka\v{c}-Moody algebra as the chiral one. 

In analogy to \eqref{eqn:flatderivative} and \eqref{eqn:vielbein}, we introduce a flat derivative $D_{\bar a}$ defined in terms of the vielbein
\begin{equation}\label{eqn:vielbeinbared}
  e^{\bar a}{}_{\bar i} = \mathcal{K}(t^a, \partial_{\bar i} \gamma\,  \gamma^{-1})
    \quad \text{as} \quad
  D_{\bar a} = e_{\bar a}{}^{\bar i} \partial_{\bar i}\,.
\end{equation}
In order to distinguish between the chiral and the anti-chiral part, it is convenient to use bared indices so that the commutator is written as
\begin{equation}
  [ D_{\bar a}, D_{\bar b} ] = F_{\bar a\bar b}{}^{\bar c} D_{\bar c}\,.
\end{equation}
In the left/right symmetric WZW model corresponding to a geometric background, the bared and unbared structure coefficients are related by
\begin{equation}\label{eqn:signfandfbar}
  F_{\bar a\bar b}{}^{\bar c} = - F_{ab}{}^c \,.
\end{equation}
However in general,  we want to treat them as independent quantities. The derivative in \eqref{eqn:vielbeinbared} acts on the right-moving coordinates $x^{\bar i}$ only. Combining these $D$ right-moving coordinates with the $D$ left-moving ones, we obtain a doubled space parameterized by the $2D$ coordinates $X^I = (x^i, x^{\bar i})$. From this world-sheet perspective it is very natural to introduce the doubled derivative $\partial_I = (\partial_i ,
\partial_{\bar i})$ and the doubled vielbein
\begin{equation}
\label{genebeinchen}
  E_A{}^I = \begin{pmatrix}
    e_a{}^i & 0 \\
    0 & e_{\bar a}{}^{\bar i}
  \end{pmatrix}
\end{equation}
giving rise to a doubled flat derivative
\begin{equation}
  D_A = E_A{}^I \partial_I
    \quad \text{with} \quad
  [D_A, D_B] = F_{AB}{}^C D_C\,.
\end{equation}
At this point, one realizes a striking similarity to the flux formulation of DFT. The latter also uses a flat doubled derivative giving rise to the same algebra (see e.g. \cite{Hohm:2010xe,Geissbuhler:2013uka}). However, the details are different, as here we are considering a CFT background, whereas in traditional DFT the doubled vielbein is introduced for fluctuations. The individual entries in the vielbein are also different, e.g. in \eqref{genebeinchen} the background B-field is sort of hidden in the left and right moving frames $e_a{}^i$ and $e_{\bar a}{}^{\bar i}$. Recall that the distinction between these two frames  only exist for a CFT in the first place.

It is straightforward to generalize the structure constants $d_{\lambda_1   q_1\,\lambda_2 q_2}$ and $C_{\lambda_1 q_1\,\lambda_2 q_2\,\lambda_3 q_3}$ to the  combination of the chiral and anti-chiral fields $\phi_R$
\begin{align}
  d_{R_1\,R_2} &= \int d^{2D} X \sqrt{|H|}\, Y_{R_1}^*\, Y_{R_2}^{} = \delta_{R_1 R_2}
    \quad \text{and} \\
  C_{R_1\,R_2\,R_3} & = \int d^{2D} X \sqrt{|H|}\, Y_{R_1}^*\, Y_{R_2}^{}\, Y_{R_3}^{}
\end{align}
with
\begin{equation}\label{eqn:metricH}
  Y_R(X^I) = y_{\lambda q}(x^i)\, \bar y_{\bar \lambda \bar q}(\bar x^{\bar i})
    \, ,\quad
  H_{IJ} = E^A{}_I E^B{}_J S_{AB} \quad \text{and} \quad
  S^{AB} = \begin{pmatrix} 
      \eta^{ab} & 0 \\
      0 & \eta^{\bar a\bar b} 
    \end{pmatrix} \,.
\end{equation}

As we will see, all expressions arising in the CSFT calculation in the next section can be eventually reduced to two different off-shell amplitudes of the primaries $\phi_R$. In the vertex notation \cite{Zwiebach:1992ie,Taylor:2003gn}, these amplitudes read
\begin{align}\label{eqn:basic2point}
  \langle {\cal R}_{12} | \phi_{R_1} \rangle_1 | \phi_{R_2} \rangle_2 &= \lim_{w_i \to 0}\langle 
    I \circ \phi_{R_1}(w_1, \bar w_1)\;
    \phi_{R_2}(w_2,\bar w_2) \rangle \quad \text{and} \\
  \label{eqn:basic3point}
  \langle {\cal V}_3 | \phi_{R_1}\rangle_1 |\phi_{R_2}\rangle_2 |\phi_{R_3}\rangle_3 &= \lim_{w_i \to 0}\langle 
    I\circ f_1 \circ \phi_{R_1}(w_1,\bar w_1)\; f_2 \circ \phi_{R_2}(w_2,\bar w_2)
    \; f_3 \circ \phi_{R_3}(w_3,\bar w_3) \rangle\,,
\end{align}
where $\langle {\cal R}_{12} |$ denote the so-called reflector and state $\langle {\cal V}_3 |$ the three-point vertex. Moreover,  $I$ is the  BPZ conjugation defined as
\begin{equation}
  I(w) = \frac{1}{w} \quad \text{and} \quad
  I\circ \phi_R(w,\bar w) = w^{-2 h_R} \bar w^{-2 \bar h_R} \phi_R(I(w), \bar I(\bar w))
\end{equation}
Furthermore, 
\begin{equation}
 f_i(w_i)=w_{i\,0} + \rho_i w_i + \mathcal{O}(w_i^2)=w
\end{equation}
is a conformal mapping between the local coordinates $w_i$ around the $i$-th puncture of the sphere $S^2$ and global coordinates $w$. We fix the punctures to $(w_{1\,0}, w_{2\,0}, w_{3\,0}) = (\infty, 0, 1)$. The parameter $\rho_i$ appearing in $f_i$ is called mapping radius \cite{Belopolsky:1994sk}. We will comment on its significance later. Note that for  Virasoro primaries, like $\phi_R$, a  conformal transformation act as
\begin{equation}
  f_i \circ \phi_R(w_i, \bar w_i) = \left(\frac{d f_i}{d w_i}\right)^{h_{R_i}}
    \left(\frac{d \bar f_i}{d \bar w_i}\right)^{\bar h_{R_i}} \phi_R(f_i(w_i),
    \bar f_i(\bar w_i))\, .
\end{equation}
 
An important consistency condition of CSFT is that all primaries have to be level matched ($h_R = \bar h_R$). In this case, the off-shell amplitudes take the simple form
\begin{align}
  \langle {\cal R}_{12} | \phi_{R_1} \rangle_1 | \phi_{R_2} \rangle_2 &= d_{R_1\,R_2}  \quad \text{and} 
    \quad \label{eqn:basicR12}\\
  \langle {\cal V}_3 | \phi_{R_1}\rangle_1 |\phi_{R_2}\rangle_2 |\phi_{R_3}\rangle_3 &= |\rho_1|^{2 h_{R_1}} 
    |\rho_2|^{2 h_{R_2}} |\rho_3|^{2 h_{R_3}} \, C_{R_1\,R_2\,R_3} \label{eqn:basicV3}\,.
\end{align}
Now, we have introduced all the necessary tools to perform the CSFT calculations in the next section.

\section{${\rm DFT}_{\rm WZW}$ action and gauge transformations from CSFT}\label{sec:CSFTaction&gauge}
After having discussed the details of the world sheet theory, the corresponding CFT correlation functions and off-shell amplitudes, we present the CSFT calculations in this section. We start with introducing the string fields describing a massless closed string state on a group manifold and the parameter for its gauge transformations. Then, from CSFT we derive the effective ${\rm DFT}_{\rm WZW}$
action and its gauge transformations up to cubic order. After introducing a version of the strong constraint, we simplify the results by applying the same field redefinitions as in \cite{Hull:2009mi}. Interestingly, the form of the strong constraint differs from the one of  DFT. Finally, we calculate the gauge algebra (C-bracket) and check its closure under the new strong constraint.

Throughout the remainder of this paper, we will work in the large level $k$ limit corresponding to the large radius limit of the group manifold. Therefore, many of the quantities we will compute receive higher order in $k^{-1}$ corrections corresponding to $\alpha'$ corrections.

\subsection{String fields for massless excitations and the weak constraint}\label{sec:stringfields}
The starting point for the CSFT calculations are two string fields $|\Psi\rangle$ and $|\Lambda\rangle$. They are level matched and in Siegel gauge \cite{Siegel:1988yz}. Thus they are annihilated by
\begin{equation}\label{eqn:levelmatching}
  L_0 - \bar L_0
    \quad \text{and} \quad
  b_0^- = b_0 - \bar b_0\,.
\end{equation}
The first one has ghost number two and the second one has ghost number one.  The general string field consists of fields corresponding to all order Ka\v{c}-Moody modes acting on the Ka\v{c}-Moody ground states $|\phi_R\rangle$. Recall that for toroidal DFT, one restricts the string field to just the lowest lying massless oscillation modes acting on the Kaluza-Klein (momentum) and winding ground states. Since in this case there does not exist a regime for the radius such that all these states are lighter than the first excited oscillation mode, this is not a low-energy truncation of the theory. However, the strong constraint prohibits simultaneous winding and momentum excitations in the same direction. In this sense, for DFT the torus can always be chosen in a way permitting a consistent low-energy truncation.

For the WZW model the situation is similar. Analogous to the toroidal case, we first remove all massive string excitations from the string field. Then, we recall the explicit Sugawara form of the Virasoro operator
\begin{equation}\label{eqn:Lm}
  L_m = -\frac{\alpha'}{4} \Bigl( 1 - h^\vee k^{-1} \Bigr) \eta^{ab} \sum\limits_n :j_{a,n-m}\; j_{b,-n}: + \mathcal{O}(k^{-3})
\end{equation}
where we have expanded the prefactor as
\begin{equation}
 - \frac{\alpha'}{2} \frac{k}{2(k + h^\vee)} = -\frac{\alpha'}{4} ( 1
 - h^\vee k^{-1} + \cdots )
\end{equation}
and have taken into account that the chiral currents $j_a$ and $j_b$ include a normalization factor  $k^{-1/2}$. Hence, we find exactly the order $\mathcal{O}(k^{-3})$ stated in \eqref{eqn:Lm}. Then, e.g. the state $j_{a,-1} \, j_{\bar b,-1}\, c_1 \bar c_1 |\phi_R\rangle$ is still present in the truncated string field and its mass is given by
\begin{equation}
\label{hannover96}
  (L_0 + \bar L_0) j_{a,-1} j_{\bar b,-1} c_1 \bar c_1 |\phi_R\rangle = \frac{1}{2 k} ( 1 - h^\vee k^{-1} ) ( c_2 (\lambda) + c_2(\bar \lambda) ) j_{a,-1} j_{\bar b,-1} c_1 \bar c_1 |\phi_R\rangle + \mathcal{O}(k^{-3})
\end{equation}
where $c_2(\lambda)$ denotes the quadratic Casimir of the representation with the highest weight $\lambda$. Now, for a fixed ground state in the representation $\lambda$, one can always choose the level $k$ large enough so that the mass in \eqref{hannover96} is much smaller than one. For fixed level $k$, there exist always ground states with a mass much larger than one\footnote{For instance for $SU(2)_k$, there are finitely many highest
  weight representations with conformal dimension $h={l(l+2)\over 4(k+2)}$ 
  with $0\le l\le k$. The state carrying  highest mass is $l=k$ with $h=k/4$.}. This is the same behavior as for the toroidal case, but only after one applies the strong constraint there. Thus, the truncated string field is given by
\begin{align}
  |\Psi\rangle &= \sum\limits_R \Bigl[ {\textstyle\frac{\alpha'}{4}} \epsilon^{a\bar
      b}(R)\, j_{a, -1}\, \bar j_{\bar b,-1}\, c_1\bar c_1 + e(R)\, c_1 c_{-1}
    + \bar e(R)\,\bar c_1 \bar c_{-1} + \nonumber \\ &
    \phantom{aaaaaaaaaaaaaaaa}
  {\textstyle \frac{\alpha'}{2}} \bigl( f^a(R)\, c_0^+c_1\, j_{a,-1} + f^{\bar b}(R)\,
  c_0^+\bar c_1\, \bar j_{\bar b,-1} \bigr) \Bigr] |\phi_R\rangle \label{eqn:stringfield} \, ,
\end{align}
and for the gauge parameters the corresponding string field is
\begin{equation}
\label{gaugestringfield}
  |\Lambda \rangle = \sum\limits_R \Bigl[ {\textstyle\frac{1}{2}} \lambda^a(R) j_{a,-1}
    c_1 - {\textstyle \frac{1}{2}} \lambda^{\bar b}(R)\, \bar j_{\bar b,-1}\, \bar c_1 +
    \mu(R)\, c_0^+ \Bigr] |\phi_R\rangle \, 
\end{equation}
with $c_0^\pm = \frac{1}{2} ( c_0 \pm \bar c_0 )$. The fields $\epsilon^{a\bar b}(R)$, $e(R)$ etc. can be considered as fluctuations around the WZW background. In contrast to the toroidal case \cite{Hull:2009mi}, in \eqref{eqn:stringfield} one does not sum over winding and momentum modes but over the different representations $R = (\lambda q\,\bar\lambda\bar q)$. 

Now, let us derive the consequences of the level-matching constraint \eqref{eqn:levelmatching} in more detail. This will guide us to the DFT$_{\rm WZW}$ generalization of the weak and strong constraint. For that purpose, let us take a closer look at a component of the string field, like e.g. $e(R)$. We assume that the group manifold $G$ is simply-connected so that the functions $Y_{R}(X)$ introduces in section~\ref{sec:fundamentalamp} form a basis for the square-integrable functions $L^2(G)$ on $G$. Hence, we are able to express each $e(X)\in L^2(G)$ as 
\begin{equation}\label{eqn:modeexpansion}
  e(X) = \sum\limits_R e(R)\, Y_{R}(X)\, .
\end{equation}
For this field, the level matching constraint \eqref{eqn:levelmatching} translates into
\begin{equation}
  \left( D_a D^a - D_{\bar a} D^{\bar a} \right) e = 0\,.
\end{equation}
This can be compactly expressed in terms of the doubled index notation introduced in section \ref{sec:fundamentalamp}. Introducing the $O(D,D)$ type constant metric
\begin{equation}\label{eqn:etaAB}
  \eta^{AB} = \begin{pmatrix} \eta^{ab} & 0 \\
    0 & -\eta^{\bar a\bar b}
  \end{pmatrix}
    \quad \text{and it's inverse} \quad
  \eta_{AB} = \begin{pmatrix} \eta_{ab} & 0 \\
    0 & -\eta_{\bar a\bar b}
  \end{pmatrix}\,,
\end{equation}
the level matching constraint reads
\begin{equation}\label{eqn:weakconstflat}
  \eta^{AB} D_A D_B \, \cdot = D_A D^A \, \cdot = 0 \,.
\end{equation}
Here, $\cdot$ stands for the physical fields $e,\,\bar e,\,\epsilon^{a\bar b},\,f^a,\,f^{\bar b}$ and the gauge parameters $\lambda^a,\,\lambda^{\bar  b},\,\mu$. In this notation, it closely resembles the weak constraint of usual DFT. However, it is given in flat and not in curved indices so that for a proper comparison, we have to transform it into curved ones. To this end, we employ the identities
\begin{equation}
  \Omega_b{}^{ba} = - \Omega_b{}^{ab} + \partial_i g^{ij} e^a{}_j
    \quad \text{with the coefficients of anholonomy} \quad
  \Omega_{ab}{}^c = e_a{}^i \partial_i e_b{}^j e^c{}_j
\end{equation}
and 
\begin{equation}
  F_{ab}{}^b = 0 = 2\Omega_{[ab]}{}^b = \Omega_{ab}{}^b - \Omega_{ba}{}^b
    \quad \Rightarrow \quad
  \Omega_{ab}{}^b = \Omega_{ba}{}^b\,,
\end{equation}
which follows from unimodularity of the Lie algebra $\mathfrak{g}$, as required in \eqref{eqn:unimodular}. Moreover, for a constant dilaton $\phi$ one gets
\begin{equation}\label{eqn:Omegaabb}
  2 D^a d = \Omega^a{}_b{}^b\,,
    \quad \text{where} \quad
  d = \phi - \frac{1}{2} \log \sqrt{|G|}
\end{equation}
is the generalized dilaton of DFT. Combining these results we obtain the relation
\begin{equation}
  \Omega_b{}^{ba} = - 2 D^a d + \partial_i g^{ij} e^a{}_j
\end{equation}
by which one finds
\begin{equation}
  D_a D^a \cdot = (\Omega_b{}^{ba} D_a + g^{ij} \partial_i \partial_j) \cdot =
   ( - 2\partial_i d\, \partial^i + \partial_i \partial^i ) \cdot \,.
\end{equation}
The analogous relation holds for bared indices, as well. Thus, with
\begin{equation}
  \eta^{IJ} = E_A{}^I E_B{}^J \eta^{AB} = \begin{pmatrix}
    g^{ij} & 0 \\
    0 & -g^{\bar i \bar j}
  \end{pmatrix}
\end{equation}
we obtain for \eqref{eqn:weakconstflat} in curved indices
\begin{equation}\label{eqn:weakconstcurved}
  ( \partial_I \partial^I - 2\, \partial_I d\, \partial^I ) \cdot = 0\,.
\end{equation}
Note that curved doubled indices are raised and lowered with $\eta^{IJ}$ which in this case is {\it not constant}. This is an essential difference to traditional DFT. It implies that one cannot pull $\eta^{IJ}$ in and out of partial derivatives so that e.g. the expressions $\partial^I \partial_I=\eta^{IJ} \partial_J \partial_I$ and $\partial_I \partial^I=\partial_I(\eta^{IJ} \partial_J)$ are not equivalent.

The weak constraint \eqref{eqn:weakconstcurved} can be further simplified by invoking the definition of a covariant derivative
\begin{equation}
  \nabla_I V^J = \partial_I V^J + \Gamma_{IK}{}^J V^K \,.
\end{equation}
In general, not all components of the generalized Christoffel symbols $\Gamma_{IK}{}^J$ are fixed but, as we will show in section \ref{sec:cbracket}, the compatibility with partial integration yields  
\begin{equation}
  \Gamma_I = \Gamma_{JI}{}^J = -2 \partial_J d \,.
\end{equation}
Hence, one can rewrite \eqref{eqn:weakconstcurved} as
\begin{empheq}[box=\fbox]{equation}\label{eqn:weakconstcov}
  \nabla_I \partial^I \cdot = 0\, .
\end{empheq}
We will also see in  section \ref{sec:cbracket} that one can require  metric compatibility $\nabla_I \eta^{JK}=0$. Using this, the expression  \eqref{eqn:weakconstcov} does not suffer from the problem $\partial_I \partial^I \cdot \ne \partial^I \partial_I \cdot$ outlined above. Indeed, it follows immediately that $\nabla_I \partial^I \cdot = \nabla^I \partial_I \cdot$\,. 

Applying \eqref{eqn:weakconstflat} to a product of two elementary objects we arrive at the strong constraint
\begin{empheq}[box=\fbox]{equation}\label{eqn:strongconst}
  D_A f\, D^A g= \partial_I f\, \partial^I g =0\, .
\end{empheq}
Note that in curved indices this constraint also involves the non-constant metric $\eta^{IJ}$.

\subsection{Action and gauge transformations}
In closed string field theory, the tree level action is given by \cite{Zwiebach:1992ie,Hull:2009mi}
\begin{equation}\label{eqn:treelevelaction}
  (2\kappa^2) S= \frac{2}{\alpha'} \Bigl( \{\Psi, Q \Psi\} + \frac{1}{3} \{\Psi, \Psi, \Psi\}_0 +
    \frac{1}{3\cdot4} \{\Psi,\Psi,\Psi,\Psi\}_0 + \dots \Bigr)
\end{equation}
where $\psi$ denotes the string field \eqref{eqn:stringfield}. It is a sum over infinitely many string vertices $\{\cdot,\,\cdots\,,\cdot\}_0$ evaluated at the genus zero world-sheet $S^2$. These are also called string functions. As in \cite{Hull:2009mi}, here we will evaluate these vertices up to order three. The fourth order term is already quite challenging as it involves an integral over a region in $\mathbb C$, whose boundary is not analytically known. First we will calculate the quadratic order and then discuss the appearance of Ward identities which will be used along the line of \cite{Rastelli:2000iu} to calculate the cubic order. This will give the simplest interactions among the components of the string field.
 
Besides the action \eqref{eqn:treelevelaction}, CSFT admits to calculate gauge transformations of the action, too. They read
\begin{equation}
\label{eqn:treelevelgauge}
  \delta_\Lambda \Psi = Q\Lambda + [\Lambda, \Psi]_0 + \frac{1}{2!} [\Lambda, \Lambda, \Psi]_0 + \dots
\end{equation}
and are parameterized by $\Lambda$, the ghost number one string field introduced in \eqref{gaugestringfield}. Here, the string product $[\cdot, \cdot]_0$ appears, which is connected to the string function by the identity
\begin{equation}
  [B_1, \dots, B_n]_0 = \sum\limits_s |\phi_s \rangle \{\phi_s^c, B_1, \dots, B_n\}_0\,.
\end{equation}
The string fields $\phi_s^c$ are called conjugate fields of $\phi_s$. Since  for CSFT on the torus, the CFT is free, it is straightforward to obtain the conjugate fields. However, on group manifolds, the world-sheet theory is in general interacting so that the notion of conjugate fields becomes more involved. We will tackle this problem while discussing the gauge transformations at quadratic order.

\subsubsection{CSFT at quadratic order}\label{sec:freetheory}
Let us start with the leading order CSFT action
\begin{equation}\label{eqn:2pointstringfunc}
  \{\Psi, Q \Psi\} = \langle\Psi| c_0^- Q | \Psi\rangle\,
\end{equation}
with the BRST operator given by\footnote{\label{foot:nilpotBRST} In a theory
  free from conformal anomalies, the
  BRST operator has to be nilpotent. This is only the case if the central
  charge $c_\mathrm{gh}=-26$ of the ghost system cancels the one of the
  bosons. Thus, we have to add $26-D$ ($D$ is the dimension compact Lie
  algebra $\mathfrak{g}$) abelian directions. Furthermore, for finite level
  $k$ we need a linear dilaton in one of the abelian directions.}
\begin{equation}
  Q = \sum\limits_m \Bigl( :c_{-m} L_m: + \frac{1}{2}:c_{-m} L^{gh}_{m}:
  \Bigr) + \text{anti-chiral}\, .
\end{equation}
We know the exact definition of $L_m$ and $L^{gh}_m$ in terms of the modes  $j_{a m}$, $c_m$ and $b_m$, but for most purposes we only need to employ the commutator
\begin{equation}\label{eqn:algebraLphi}
  [L_m, \phi_n] = \Big((h - 1) m - n\Big)
\end{equation}
between a Virasoro generator and a primary field $\phi$ of conformal weight and similarly for the ghost contribution $L^{gh}_m$ $h$. 

As we have already defined in \eqref{eqn:basicR12}, a convenient way to express the expectation value \eqref{eqn:2pointstringfunc} is in terms of the reflector state $\langle {\cal R}_{12}|$, namely
\begin{equation}\label{eqn:actionR12}
  \langle \Psi | c_0^- Q | \Psi\rangle = \langle {\cal R}_{12} | \Psi \rangle_1 c_0^{-(2)} Q^{(2)} | \Psi\rangle_2\,.
\end{equation}
Then, we can use the identities \cite{Zwiebach:1992ie}
\begin{equation}\label{eqn:wardR12}
  \langle {\cal R}_{12} | c_m^{(1)} + c_{-m}^{(2)} = 0 \quad\text{and}\quad
  \langle {\cal R}_{12} | j_{a,m}^{(1)} + j_{a,-m}^{(2)} = 0 
\end{equation}
to move operators from one side of the reflector to the other. As \eqref{eqn:actionR12} is bilinear, one can treat each term in the string field \eqref{eqn:stringfield} separately. To continue, we use the following algorithm: On each side of the reflector state we move operators annihilating the primary $|\phi_R\rangle$ or the ghost vacuum to the right by using the commutation relations \eqref{eqn:algebraLphi} and \eqref{eqn:algebrajj}. This procedure is called normal ordering. It is performed in such a way that the Virasoro generators are transported directly to the primary field in each slot of the reflector state. Only $L_0$ and $L_{-1}$ survive this procedure. According to \eqref{eqn:Lm}, one can replace $L_0$ and $L_{-1}$ by
\begin{align}
  L_0 |\phi_R\rangle &= -\frac{\alpha'}{4} (1 - h^\vee k^{-1} + \dots )
  \eta^{ab}\, j_{a,0}\, j_{b,0} |\phi_R\rangle\,, \nonumber \\
    L_{-1} |\phi_R\rangle &= -\frac{\alpha'}{2} (1 - h^\vee k^{-1} + \dots) \eta^{ab}\, j_{a,-1}\, j_{b,0} |\phi_R\rangle\,
    \label{eqn:L0L1alphapexpansion}
\end{align}
for large $k$. Afterwards, we perform normal ordering again until only zero modes or creation operators are left over. All operators acting on the first part of $\langle {\cal R}_{12}|$ are moved to the second one utilizing the identities \eqref{eqn:wardR12}. We establish normal ordering and so that, finally, only zero modes are left over.

Just to give an impression,  one of the many terms  of the resulting expression is
\begin{equation}
  \{\Psi, Q \Psi\} = \dots + \frac{\alpha'}{2} \sum\limits_{R_1,\, R_2} \bar e(R_1)\, e(R_2)\, \eta^{ab}\,
    \langle {\cal R}_{12}| \phi_{R_1}\rangle_1 c_{-1}\bar c_{-1} c_0 c_1 \bar c_1\,
    j_{a,0}\, j_{b,0} |\phi_{R_2}\rangle_2 + \dots \,.
\end{equation}
To get rid of the ghost zero modes $c_{-1}$, $c_0$ and $c_1$ we apply the ghost overlap\footnote{We use the convention of \cite{Hull:2009mi} which   differs by a sign from the earlier works like \cite{Zwiebach:1992ie}.}
\begin{equation}\label{eqn:ghostoverlap}
  \langle \phi_{R_1} | c_{-1}\bar c_{-1} c_0^- c_0^+ c_1 \bar c_1 |\phi_{R_2} \rangle := \delta_{R_1\, R_2}
    \qquad \Leftrightarrow \qquad
  \langle \phi_{R_1} | c_{-1} c_0 c_1 \bar c_{-1} \bar c_0 \bar c_1 |
  \phi_{R_2} \rangle = 2 \delta_{R_1\,R_2}\, .
\end{equation}
Recalling the two-point amplitude \eqref{eqn:basicR12} and combining it with the substitution 
\begin{equation}
  j_{a,0} |\phi_R\rangle = t_a |\phi_R\rangle \qquad \text{and} \qquad
    t_a \rightarrow D_a\,,
\end{equation}
we obtain the final result
\begin{equation}
 (2\kappa^2)S= \dots + \frac{\alpha'}{2} \int d^{2D} X \,\sqrt{|H|} \, \bar e\,  D_a D^a e + \dots \,.
\end{equation}
After a tedious computation, at leading order $\mathcal{O}(k^{-1})$ the complete quadratic action reads
\begin{align}
  (2\kappa^2) S^{(2,-1)} &= \int d^{2D}\, \sqrt{|H|}\, \Bigl[ {\textstyle \frac{1}{4}} \epsilon_{ab} \square \epsilon^{ab} + 2\, \bar e \,\square e - f_a\, f^a - f_{\bar b}\, f^{\bar b} \nonumber \\
  &\qquad\qquad - f_a ( D_{\bar b} e^{a\bar b} - 2 D^a \bar e) + f_{\bar b} ( D_a e^{a\bar b} + 2 D^{\bar b} e) \Bigr] \label{eqn:S2-1}
\end{align}
where the generalized Laplace operator is defined as
\begin{equation}
  \square = \frac{1}{2} \left( D_a D^a + D_{\bar a} D^{\bar a} \right)\, .
\end{equation}
Let us make a couple of comments:
\begin{itemize}
  \item Note that we assumed the auxiliary fields $f_a$ and
    $f_{\bar a}$ to be proportional to $k^{-1/2}$, as otherwise we would 
    also find additional terms in \eqref{eqn:S2-1}. This situation is in
    total accordance with toroidal DFT, where the auxiliary fields are
    also weighted by an additional factor $\sqrt{\alpha'}$.
  \item On the torus, the vielbein $E_A{}^I$ is independent of the
    coordinates $X^I$, so that one can simply substitute the flat 
    coordinates in \eqref{eqn:S2-1} by curved ones. In this way, one 
    exactly reproduces the result derived in  \cite{Hull:2009mi}. 
  \item Even though \eqref{eqn:S2-1} looks like the one for toroidal
    DFT, there is a substantial difference in that the derivatives
    appearing there do not commute.
\end{itemize}
At subleading orders in $k^{-1}$ the difference become even more striking. Recall that such corrections have the interpretation of $\alpha'$ corrections.Whereas for the toroidal case such corrections are absent in the CFT action at quadratic order, for the WZW model there exist a whole series of them. Thus, all quantities on the world-sheet receive corrections which is already reflected in \eqref{eqn:L0L1alphapexpansion}, where the Virasoro generators $L_0$ and $L_{-1}$ receive corrections in all orders of $k^{-1}$.

\vspace{0.2cm}
Now, we come to the evaluation of the gauge transformation \eqref{eqn:treelevelgauge} at second order, which involve the conjugate fields $\phi_s$. These are defined by the relation
\begin{equation}\label{eqn:defconjugatestate}
  \{\phi_s^c, \phi_{s'}\}_0 = \langle \phi_s^c| c_0^- |\phi_{s'} \rangle =
  \langle {\cal R}_{12} | \phi_s^c\rangle_1 c_0^{-(2)} |\phi_{s'}\rangle_2 = \delta_{s s'}\,.
\end{equation}
Since $j_{a,-1}$ and $j_{\bar b,-1}$ are the only creation operators appearing in the massless string fields, it is sufficient to know the conjugate field of $\phi_s = j_{a,-1} |\phi_R\rangle$ with $s=(a,R)$ (and its anti-chiral counterpart). A first guess for this conjugate field is $\phi_s^c = j^{a}_{-1} |\phi_R\rangle$, which  is along the lines of the abelian case. Evaluating \eqref{eqn:defconjugatestate}, we obtain
\begin{equation}\label{eqn:defconjsf}
  \langle {\cal R}_{12} | j_{-1}^{a\,(1)} |\phi_{R_1}\rangle_1\, j_{b,-1}^{(2)} |\phi_{R_2}\rangle_2 =
    -F^a{}_b{}^c \, \langle {\cal R}_{12} | \phi_{R_1}\rangle_1\, j_{c,0}^{(2)} |\phi_{R_2}\rangle_2 
    + \frac{2}{\alpha'}\, \delta^a_b\, \delta_{R_1\,R_2} \,.
\end{equation}
We realize that, even though the second term on the right hand side looks quite good, the first one spoils everything. We can get rid of this term by instead defining the conjugate field as
\begin{equation}\label{eqn:conjsf}
  \phi_s^c = \Bigl( {\textstyle \frac{\alpha'}{2}} - {\textstyle \left( \frac{\alpha'}{2} \right)^{3/2}} k^{-1} \Bigr)
    \Bigl( j^a_{-1} + {\textstyle \frac{\alpha'}{2}}\, F^{abc}\, j_{c,0}\, j_{b,-1} \Bigr)\, .
\end{equation}
Indeed, after some algebra and using \eqref{eqn:defetaab}, up to order $k^{-1}$, this ansatz gives rise to the desired result 
\begin{equation}\label{eqn:conjja-1}
  \langle {\cal R}_{12} | \phi_{s_1}^c \rangle_1 j_{b,-1}^{(2)} |\phi_{R_2}\rangle_2 =
    \delta^a_b\, \delta_{R_1\,R_2}  + \mathcal{O}(k^{-3/2})\,,
\end{equation}
which  is an improvement in comparison to our first guess. There it was only satisfied up to the order $k^{-1/2}$. In general, one has to determine the conjugate fields order by order in inverse powers of $k$. However, for all orders we are considering in this paper, \eqref{eqn:conjja-1} is sufficient.

Now, we have collected all ingredients to calculate the gauge transformations
\begin{equation}
  \delta_\Lambda \Psi = \sum\limits_s |\phi_s \rangle \{\phi_s^c, Q \Lambda \}_0\,,
\end{equation}
using the same techniques as for computing the CSFT action. In the end, at leading order $\mathcal{O}(k^{-1})$ we obtain the gauge transformations
\begin{align}
  \delta_\Lambda \epsilon_{a\bar b} &= D_a \lambda_{\bar b} + D_{\bar b} \lambda_a &
  \delta_\Lambda e      &= \mu - \frac{1}{2} D_a \lambda^a &
  \delta_\Lambda f_a    &= D_a \mu - \frac{1}{2} \square \lambda_a \\
  &&
  \delta_\Lambda \bar e  &= \mu + \frac{1}{2} D_{\bar b} \lambda^{\bar b} &
  \delta_\Lambda f_{\bar b}  &= D_{\bar b}\, \mu + \frac{1}{2} \square \lambda_{\bar b}\,.
\end{align}
These and the quadratic action \eqref{eqn:S2-1} possess the $\mathds{Z}_2$ symmetry
\begin{equation}\label{eqn:Z2sym}
  \epsilon_{a\bar b} \,\, \leftrightarrow \,\, \epsilon_{\bar b a}\,, \quad
  D_a \,\, \leftrightarrow \,\, D_{\bar a}\,, \quad
  f_a \,\, \leftrightarrow \,\, - f_{\bar a}\,, \quad
  e \,\, \leftrightarrow \,\, - \bar e\,, \quad
  \lambda_a \,\, \leftrightarrow \,\, \lambda_{\bar a}
    \quad \text{and} \quad
  \mu \,\, \leftrightarrow \,\, -\mu\,,
\end{equation}
which  is a direct consequence of vanishing (anti-)commutators between chiral and anti-chiral operators in the theory. 

\subsubsection{Interactions at cubic order}\label{sec:cubicorder}
In this section we compute the string function 
\begin{equation}
\label{stringfuncthree}
  \{\Psi, \Psi, \Psi\} = \langle {\cal V}_3 | \Psi \rangle_1 |\Psi \rangle_2 |\Psi \rangle_3 \,,
\end{equation}
which forms the cubic part of the tree-level action \eqref{eqn:treelevelaction}. Even though \cite{Rastelli:2000iu} considers open string field theory, our closed CSFT computation is very analogous.

From the discussion in section~\ref{sec:worldsheet}, we know that each mode $j_{a,n}$ of the current $j_a(z)$ is a symmetry generator of our theory. Hence, the variation
\begin{equation}\label{eqn:jsymm}
  \delta_\varepsilon \langle f_1 \circ V_1\; f_2 \circ V_2\; f_3 \circ V_3 \rangle =
  \oint \frac{d z}{2\pi i} \langle \varepsilon(z) j_a(z) I \circ f_1 \circ V_1\; f_2 \circ V_2\; f_3 \circ V_3 \rangle = 0
\end{equation}
has to vanish for arbitrary vertex operators $V_i$. In the vertex $\langle V_3|$ notation introduced in \eqref{eqn:basic3point}, this expression translates into \cite{Rastelli:2000iu}
\begin{equation}\label{eqn:vertexrulegen}
  \sum\limits_{i=1}^3 \oint_{\mathcal{C}_i} \frac{dz}{2\pi i}\, \langle {\cal V}_3| \varepsilon(z)\, j_a(z) = 0\,.
\end{equation}
Here, we do not explicitly write the right hand side of the equation, because it holds for arbitrary $V_i$. The integral in \eqref{eqn:jsymm} receives only contributions around the punctures introduced by the vertex operators. These punctures are enclosed by the contours $\mathcal{C}_i$\,. To pull the integration directly in front of the corresponding vertex operator, one has to change the integration variable from $z$ to $z_i = f^{-1}_i(z)$. Since $j_a(z)$ has conformal weight one, this transformation gives rise to
\begin{equation}
  dz\, \varepsilon(z)\, j_a(z) = dz_i\, \frac{d z}{d z_i} \left( \frac{d z_i}{d z} \right)^1 \varepsilon(f_i(z_i))\, j_a(z_i) = dz_i\, \varepsilon_i(z_i)\, j_a(z_i) 
\end{equation}
with $\varepsilon_i(z_i) = \varepsilon(f_i(z_i))$. Thus, for \eqref{eqn:vertexrulegen} we obtain
\begin{equation}
  \sum\limits_{i=1}^3 \oint_{\mathcal{C}_i} \frac{d z_i}{2\pi i}\, \langle {\cal
    V}_3| \varepsilon_i(z_i)\, j_a(z_i) = 0 \,.
\end{equation}
The functions $z=f_i(z_i)$ map the local coordinates around the punctures at $z_{0\,i}=\{\infty, 0, 1\}$ to a common coordinate system $z$. In doing so, they describe the world-sheet geometry of the three-point interaction. As shown in more detail in appendix~\ref{app:3stringvertex}, they are given by
\begin{align} \label{eqn:f_2expansion}
  f_2(z_2) &= \rho z_2 + d_1 (\rho z_2)^2 + d_2 (\rho z_2)^3 + \dots\,, \quad \\
  f_3(z_3) &= \frac{1}{1 - f_2(z_3)} \quad \text{and} \quad
  f_1(z_1) = 1 - \frac{1}{f_2(z_1)}
\end{align}
with the constants
\begin{equation} \label{eqn:coeff_2expansion}
  \rho = - \frac{4}{3\sqrt{3}}\,, \qquad
  d_1  = - 1/2\, \quad \text{and} \qquad
  d_2  = - 1/16\,.
\end{equation}
Choosing $\varepsilon(z) = \rho / z$ and utilizing the mode expansion of the chiral current $j_a(z_i)$ in \eqref{eqn:modeexpja}, we obtain the Ward identity
\begin{equation}\label{eqn:wardj-1}
  \langle {\cal V}_3| \bigl( \rho \, j_{a,0}^{(1)} - \rho^2 \,j_{a,0}^{(1)} +j_{a,-1}^{(2)}  - \rho d_1\, j_{a, 0}^{(2)} +  \rho^2 (d_1^2 - d_2^2)\, j_{a, 1}^{(2)}
    - \rho^2\, j_{a,1}^{(3)}  + \dots \bigr) = 0\,.
\end{equation}
A similar argument holds for the $c$-ghosts, which are Virasoro primaries of conformal weight $-1$. Thus, the main difference is the transformation 
behavior of
\begin{align}
  dz\, \phi(z)\, c(z) & = d z_i \frac{d z}{d z_i} \left(\frac{d z_i}{d z}\right)^{-1} \phi(f(z_i))\, c(z_i) =
    d z_i\, \phi_i(z_i)\, c(z_i) 
\end{align}
with $\phi_i(z_i) = ( f'(z_i) )^{-2} \, \phi( f (z_i) )$. Again, for the specific choices
\begin{equation}
  \phi(z) = \frac{1}{(1-z) z^2} \quad \text{and} \quad
  \phi(z) = \frac{(z-2)\rho}{2(z-1)z^3}\,
\end{equation}
the two Ward identities
\begin{align}
  \langle {\cal V}_3| \bigl( \rho c_1^{(1)}+ c_0^{(2)} + \rho( 1 + 2 d_1)\, c_1^{(2)} - \rho\, c_1^{(3)} \bigr) &= 0 
    \label{eqn:wardc0} \\[0.2cm]
  \Big\langle {\cal V}_3\Big| \Bigl(  -\frac{\rho^2}{2}\, c_1^{(1)} + c_{-1}^{(2)} +
  \frac{\rho}{2} (1 + 2 d_1)\, c_0^{(2)} + \frac{\rho^2}{2}( 1 + 2 d_1 - 4 d_1^2
  + 6 d_2 )\, c_{1}^{(2)} -\frac{\rho^2}{2}\, c_1^{(3)}  \Bigr) &= 0
    \label{eqn:wardc-1}
\end{align}
follow. For bared operators, analogous Ward identities hold.

Equipped with these Ward identities, we one can now proceed and compute the string function \eqref{stringfuncthree}. Like for the quadratic term, we again use the bilinearity of the string function and obtain $5^3=125$ different terms to calculate. Considering also their symmetries, it is sufficient to calculate only $35$ different terms  and weight them with the corresponding combinatoric prefactors.

To evaluate each of these 35 remaining string functions, we apply the following algorithm: First we use one of the Ward identities \eqref{eqn:wardj-1}, \eqref{eqn:wardc0} or \eqref{eqn:wardc-1} to remove the corresponding operator from the second slot of $\langle {\cal V}_3|$. Afterwards we establish normal ordering of all slots and remove terms where annihilation operators hit the primaries. We repeat this procedure until slot two of $\langle {\cal V}_3|$ contains the operators $c_1$, $\bar c_1$ and $j_{a, 0}$ only. Now, we rotate the vertex according to the rule
\begin{equation}
  \langle {\cal V}_3| V_1\rangle_1\, V_2\rangle_2\, V_3\rangle_3 = (-)^{V_1(V_2 +
    V_3)} \langle {\cal V}_3| V_2 \rangle_1\, V_3 \rangle_2\, V_1 \rangle_3
\end{equation}
and start over again by applying the Ward identities and normal ordering. Then we rotate again and we continue until all slots of $\langle {\cal V}_3|$ contain $c_1$, $\bar c_1$, $j_{a, 0}$ and $\bar j_{\bar a, 0}$ operators only. Finally, we apply the ghost overlap \eqref{eqn:ghostoverlap} giving rise to the substitution rule
\begin{equation}
  \langle {\cal V}_3 | c_1^{(1)} \bar c_1^{(1)} c_1^{(2)} \bar c_1^{(2)}
  c_1^{(3)} \bar c_1^{(3)} = \frac{2}{|\rho|^6} \langle  {\cal V}_3|
\end{equation}
where the $|\rho|^6$ term in the denominator arises because we have 6 ghosts with conformal weight $-1$. It is canceled completely by the $|\rho|^6$ due to the successive application of the Ward identities. After all these steps, only the fundamental three-point off-shell amplitudes \eqref{eqn:basicV3} are left over. Writing them in terms of an integral over the doubled space,we have to take care of the $|\rho|^{2 h_i}$ factors in \eqref{eqn:basicV3}. However, they can be expressed as
\begin{equation}
  |\rho|^{2 h_R} = |\rho|^{-\frac{\alpha'}{2 (k + h^\vee)} \square} = 1 - \frac{\alpha'}{2} \ln |\rho| \square + \dots  = 1 + \mathcal{O}(k^{-1})\,
\end{equation}
and therefore, at leading order,  do not give any contribution to the action\footnote{Even though the algorithm presented here is 
  straightforward, the calculations are lengthy and cumbersome. For that 
  purpose we developed a Mathematica package that was inspired to some 
  extent by Lambda \cite{Ekstrand:2010bp}, a package to evaluate operator 
  product expansions in vertex algebras. It also extensively uses MathGR
  \cite{Wang:2013mea} to simplify tensor expressions.}.
Finally, at leading order $O(k^{-1})$, the cubic part of the action can be
expressed as
\begin{align}
  (2\kappa^2) S^{(3,-1)} &= \int d^{2D} X \sqrt{|H|} \, \biggl[ -\frac{1}{8} \epsilon_{a\bar b} \Bigl( 
    - D_c \epsilon^{c \bar b}\, D_{\bar d} \epsilon^{a\bar d} - D_c
    \epsilon^{c \bar d} \, D_{\bar d} 
    \epsilon^{a\bar b} -  2 D^a \epsilon_{c \bar d}\, D^{\bar b} \epsilon^{c\bar d} \nonumber\\
  & \hspace{11em} + 2 D^a \epsilon_{c\bar d} \, D^{\bar d} \epsilon^{c\bar b} + 
    2 D^c \epsilon^{ad} \, D^{\bar b} \epsilon_{c \bar d} \Bigr) \nonumber\\
  & \quad - \frac{1}{4} \epsilon_{a\bar b} \Bigl( 
    F^a{}_{cd}\, \epsilon^{c\bar e}\,  D_{\bar e} \epsilon^{d \bar b} + 
    F^{\bar b}{}_{\bar c \bar d}\, \epsilon^{e \bar c}\, D_e \epsilon^{a \bar d}\Bigr) 
    -\frac{1}{12} F_{ace} \, F_{\bar d\bar b\bar f} \, \epsilon^{a\bar b}\, \epsilon^{c\bar d}\, 
    \epsilon^{e\bar f} \label{eqn:newcubic} \\
  & \quad + \frac{1}{2} \epsilon_{a \bar b}\, f^a f^{\bar b}
    -\frac{1}{2} f_a f^a \, \bar e +  \frac{1}{2} f_{\bar a} f^{\bar a}\, e \nonumber\\
  & \quad -\frac{1}{8} \epsilon_{a\bar b}\, \Bigl( D^a D^{\bar b} e\; \bar e -
    D^a e \, D^{\bar b} \bar e
    -D^{\bar b} e\,  D^a \bar e + e\, D^a D^{\bar b} \bar e \Bigr) \nonumber \\
  & \quad -\frac{1}{4} f^a \Bigl( 2 \epsilon_{a\bar b}\, D^{\bar b} \bar e + 
    D^{\bar b} \epsilon_{a\bar b}\; \bar e \Bigr)
    +\frac{1}{4} f^a \Bigl( D_a e \; \bar e - e\,  D_a \bar e \Bigr) \nonumber\\
  & \quad -\frac{1}{4} f^{\bar b} \Bigl( 2 \epsilon_{a \bar b}\,  D^a e + D^a \epsilon_{a\bar b}\; e\Bigr)
    +\frac{1}{4} f^{\bar b} \Bigl( D_{\bar b} e\; \bar e - e\, D_{\bar b} \bar e \Bigr) \biggr]\,.
\end{align}
Like already observed for the second order action \eqref{eqn:S2-1}, large parts of it resemble the original result obtained by Hull and Zwiebach. However, there are also additional terms \eqref{eqn:newcubic}, linear and quadratic in the structure coefficients $F_{abc}$. On the abelian torus they vanish and then the action \eqref{eqn:newcubic} reduces to the one derived in \cite{Hull:2009mi}. Whereas in toroidal DFT, there are kinetic terms in the action only, one of the additional terms \eqref{eqn:newcubic} represents a potential
\begin{equation}
  V = -\frac{1}{12}\, F_{ace}\, F_{\bar b\bar d\bar f}\, \epsilon^{a\bar b}\, \epsilon^{c\bar d}\, \epsilon^{e\bar f} 
\end{equation}
for the fluctuations $\epsilon_{a\bar b}$. 

In order to evaluate the gauge transformations in cubic order, we again use the conjugated string fields $\phi_s^c$ from section~\ref{sec:freetheory}. They allow to express the string product
\begin{equation}
  [\Psi, \Lambda]_0 = \sum\limits_s |\phi_s\rangle \{\phi_s^c, \Psi, \Lambda\}_0
\end{equation}
in terms of string functions, which we compute like those appearing in the action. One finally obtains for the gauge variations of the fluctuations
\begin{align}\label{eqn:deltalambdaepsilon}
  \delta_\lambda \epsilon_{a \bar b} &= - \frac{1}{4} \Bigl( 
    \lambda^c\, D_a \epsilon_{c\bar b} - D_a \lambda^c\; \epsilon_{c\bar b} + \lambda_a\, D^c \epsilon_{c \bar b}
    + 2 D^c \lambda_a\; \epsilon_{c\bar b} - \lambda_c\, D^c \epsilon_{a\bar b} 
    - 2 \lambda_c\, D^c \epsilon_{a\bar b} \Bigr) \nonumber \\
  & \quad - \frac{1}{4}\Bigl( \lambda_a\, D_{\bar b} \bar e - D_{\bar b} \lambda_a\; \bar e \Bigr)
    + \frac{1}{2}\lambda_a\, f_{\bar b} + \frac{1}{2} F_{ac}{}^d\, \lambda^c\, \epsilon_{d \bar b} \\
  \delta_\lambda e &= -\frac{1}{4} f^a\, \lambda_a + \frac{1}{8} e\, D^a \lambda_a + \frac{1}{4} \lambda_a\, D^a e \\
  \delta_\lambda \bar e &= \frac{1}{16} \bar e\, D^a \lambda_a + \frac{1}{8} \lambda_a\, D^a \bar e\,. 
\end{align}
The corresponding ones for $\lambda_{\bar a}$ arise after applying the $\mathds{Z}_2$ symmetry \eqref{eqn:Z2sym}. Here, we are not interested in the gauge transformations of the auxiliary fields $f_a$ and $f_{\bar a}$, because they are eliminated by their equations of motion in the next subsection anyway. A $\mu$-type gauge transformation acts as
\begin{equation}\label{eqn:deltamu}
  \delta_\mu \epsilon_{a\bar b} = 0\,, \quad
  \delta_\mu e = - \frac{3}{8} \mu e
    \quad \text{and} \quad
  \delta_\mu \bar e = \frac{3}{8} \mu \bar e \,.
\end{equation}

\subsection{Simplifying the action and gauge transformations}\label{sec:simpactiongauge}
Following \cite{Hull:2009mi,Hohm:2014xsa}, we simplify the action by first
fixing the $\mu$ gauge in such a way that
\begin{equation}
  e = d \quad \text{and} \quad \bar e = -d \,.
\end{equation}
Afterwards, we redefine the fields 
\begin{equation}\label{eqn:fieldredefepsilond}
  \epsilon_{a\bar b}' = \epsilon_{a\bar b} + \epsilon_{a\bar b}\, d\,, \quad
  d' = d + \frac{1}{32} \epsilon_{a\bar b}\, \epsilon^{a\bar b} \,
\end{equation}
and the gauge parameter
\begin{equation}\label{eqn:fieldredefgaugepara}
  \lambda_a' = \lambda_a + \frac{3}{4}\lambda_a\, d - \frac{1}{4}
  \lambda^{\bar b} \, \epsilon_{a \bar b}\,. 
\end{equation}
Let us briefly discuss how the level matching condition works for these redefined fields. We know that the unprimed fields in \eqref{eqn:fieldredefepsilond} have to satisfy the weak constraint \eqref{eqn:weakconstflat}. Since the primed ones contain products of unprimed fields, they do not automatically satisfy it. However, requiring also the strong constraint \eqref{eqn:strongconst} guarantees that the primed fields do it. Therefore, already at the level of this field redefinition the strong constraint is necessary.

Now, plugging the redefined quantities into the quadratic and cubic gauge
transformations and removing all contributions that are not linear in the
parameter $\lambda$ or the fields, we obtain
\begin{equation}
  \label{eqn:gaugetrafo}
  \fbox{$
  \begin{aligned}
  \delta_{\lambda} \epsilon_{a\bar b} =& D_{\bar b} \lambda_a  + \frac{1}{2}\Bigl( 
    D_a \lambda^c\, \epsilon_{c\bar b} - D^c \lambda_a\; \epsilon_{c\bar b} + \lambda_c \,D^c \epsilon_{a \bar b}
    + F_{ac}{}^d\, \lambda^c\, \epsilon_{d\bar b} \Bigr)  \\ 
  &  D_a \lambda_{\bar b} + \frac{1}{2}\Bigl( D_{\bar b} \lambda^{\bar c} \,\epsilon_{a\bar c} - 
    D^{\bar c} \lambda_{\bar b}\; \epsilon_{a \bar c} + \lambda_{\bar c} \,D^{\bar c} \epsilon_{a \bar b}
    + F_{\bar b\bar c}{}^{\bar d}\, \lambda^{\bar c}\, \epsilon_{a \bar d}
    \Bigr) \, ,\\
    \delta_\lambda d =& -\frac{1}{4} D_a \lambda^a + \frac{1}{2} \lambda_a\, D^a d - 
    \frac{1}{4} D_{\bar a} \lambda^{\bar a} + \frac{1}{2} \lambda_{\bar a}\,
    D^{\bar a} d\,,
  \end{aligned}$}
\end{equation}
where for simplicity of the notation we dropped  the prime. Except for the flux term, they have the same form as the gauge transformations of toroidal DFT. 

As already mentioned above, it is convenient to simplify the action by eliminating the auxiliary fields $f_a$ and $f_{\bar a}$. To this end, we solve their equations of motion up to quadratic order in the remaining fields, yielding
\begin{align}
  f^a & = -\frac{1}{2} D_{\bar b} \epsilon^{a \bar b} - D^a d
    +\frac{1}{2}\Bigl( \epsilon^{a \bar b} \, D_{\bar b} d + d\, D^a d\Bigr) 
    +\frac{1}{8}\Bigl(D^c \epsilon_{c\bar b}\; \epsilon^{a\bar b} - d\, D_{\bar b} \epsilon^{a \bar b} \Bigr) \\
  f^{\bar b} &=  \frac{1}{2} D_{a} \epsilon^{a \bar b} + D^{\bar b} d
    -\frac{1}{2}\Bigl( \epsilon^{a \bar b}\, D_{a} d + d\, D^{\bar b} d\Bigr) 
    -\frac{1}{8}\Bigl(D^c \epsilon_{a \bar c}\; \epsilon^{a\bar b} - d\, D_{a} \epsilon^{a \bar b} \Bigr)\,.
\end{align}
Furthermore, we apply the field redefinitions \eqref{eqn:fieldredefepsilond} which we already used to simplify the gauge transformations so that finally we obtain
\begin{equation}
  \fbox{$
  \begin{aligned}
  (2 \kappa^2)&S = \int d^{2D} X\sqrt{|H|}\, \biggl[ 
      \frac{1}{4} \epsilon_{a\bar b} \,\square \epsilon^{a\bar b} + \frac{1}{4} (D^{\bar b} \epsilon_{a\bar b})^2
      + \frac{1}{4} (D^a \epsilon_{a \bar b})^2 - 2 d\, D^a D^{\bar b} \epsilon_{a\bar b} 
      - 4 d\, \square d \\[0.1cm]
    & + \frac{1}{4} \epsilon_{a\bar b} \Bigl( D^a \epsilon_{c\bar d}\, D^{\bar b} \epsilon^{c\bar d} - 
      D^a \epsilon_{c \bar d}\, D^{\bar d} \epsilon^{c\bar b} - D^c \epsilon^{a\bar d}\, D^{\bar b} 
      \epsilon_{c\bar d} \Bigr) \\[0.1cm]
    & - \frac{1}{4} \epsilon_{a\bar b} \Bigl( F^{ac}{}_d\, D^{\bar e} \epsilon^{d\bar b} \;\epsilon_{c\bar e} 
      + F^{\bar b\bar c}{}_{\bar d}\, D^e \epsilon^{a\bar d}\; \epsilon_{e\bar c} \Bigr)
      -\frac{1}{12} F^{ace}\, F^{\bar b\bar d\bar f}\, \epsilon_{a\bar b}\, \epsilon_{c\bar d}\,
      \epsilon_{e\bar f} \\[0.1cm]
    & +\frac{1}{2} d \Bigl( (D^a \epsilon_{a\bar b})^2 + (D^{\bar b} \epsilon_{a\bar b})^2 + 
      \frac{1}{2} (D_c \epsilon_{a\bar b})^2 + \frac{1}{2} (D_{\bar c} \epsilon_{a\bar b})^2 +
      2 \epsilon^{a\bar b} ( D_a D^c \epsilon_{c\bar b} + D_{\bar b} D^{\bar c} \epsilon_{a\bar c} )
      \Bigr) \\[0.1cm]
    & +4 \epsilon_{a\bar b}\, d\, D^a D^{\bar b} d + 4 d^2\, \square d \biggr]
  \end{aligned}$}
\end{equation}
where we defined e.g.
\begin{equation}
  (D^{\bar b} \epsilon_{a\bar b})^2=
   (D^{\bar b} \epsilon_{a\bar b})(D_{\bar c} \epsilon^{a\bar c})\, .
\end{equation}
Thus, we have derived the leading order form of the DFT$_{\rm WZW}$ action,which reduces to the form of the usual  DFT action for a flat torus, containing, though, additional terms which go beyond it. First, the derivatives $D_a$ are non-commuting and, second, the fluxes $F_{abc}$ appear explicitly. 

\section{Generalized Lie derivative and C-bracket}\label{sec:cbracket}
In this section, we analyze the obtained action and gauge transformations further, focusing in particular on the generalization of some of the salient features of DFT, like the Lie derivative, the generalized metric, the C-bracket and its closure. Recall that in DFT, the latter is closely related to the implementation of the strong constraint.

To simplify the gauge transformations \eqref{eqn:gaugetrafo}, we change to doubled index notation introduced in section~\ref{sec:fundamentalamp}. Hence, we define the doubled parameter of the gauge transformations and the doubled derivative as
\begin{equation}
\label{conventa}
  \lambda^A = (\lambda^a, \lambda^{\bar a})\, ,\qquad\qquad
  D_A=(\partial_a, \partial_{\bar a})\,.
\end{equation}
As in section~\ref{sec:stringfields}, capital indices are raised and lowered with the tangent space metric $\eta$ defined in \eqref{eqn:etaAB}. Following this prescription we obtain
\begin{equation}
\label{conventb}
  \lambda_A = (\lambda_a, -\lambda_{\bar a})\, ,\qquad\qquad
  D^A = (\partial^a, - \partial^{\bar a})\,.
\end{equation}
Similarly, the structure constants in capital indices are defined as
\begin{equation}
\label{conventc}
  F_{AB}{}^C = \begin{cases}
    F_{ab}{}^c & \\
    F_{\bar a\bar b}{}^{\bar c} & \\
    0 & \text{otherwise}
  \end{cases}
  \quad \text{which e.g. gives rise to} \quad
  F_{ABC} = \begin{cases}
    F_{abc} & \\
    -F_{\bar a\bar b\bar c} & \\
    0 & \text{otherwise\,.}
  \end{cases}
\end{equation}
In the remainder of this section, these conventions will be often used.

\subsection{Generalized Lie derivative and metric}
\label{sec:generalized metric}
Now, we want to see whether the gauge transformations \eqref{eqn:gaugetrafo} encode the notation of a generalized Lie derivative. The non-trivial issue is that the right hand side of \eqref{eqn:gaugetrafo} is given in terms of an expansion up to linear order in the small fluctuation $\epsilon_{a\bar b}$. Therefore, we first have to ``integrate'' this relation, which we do
following the procedure outlined for the generalized metric formulation of 
DFT in \cite{Hohm:2010jy,Hohm:2010pp}.

For that purpose, consider first the symmetric transformation
$\mathcal{H}^{AB}$  leaving $\eta$  invariant
\begin{equation}
  \mathcal{H}^{AC} \eta_{CD} \mathcal{H}^{DB} = \eta^{AB}\,.
\end{equation}
A simple example for such a matrix is $S^{AB}$. A small perturbation of it,which is still compatible with the properties of  $\mathcal{H}^{AB}$, is called $\epsilon^{AB}$. Therefore, $\epsilon^{AB}$ has to be symmetric and has to satisfy the relation
\begin{equation}\label{eqn:O(D,D)gens}
  \epsilon^{AC} \eta_{CD} S^{DB} + S^{AC}\eta_{CD} \epsilon^{DB} +
  \mathcal{O}(\epsilon^2) = 0\, .
\end{equation}
The most general, symmetric solution for this equation reads
\begin{equation}
  \epsilon^{AB} = \begin{pmatrix}
    0 & - \epsilon^{a\bar b} \\
    - \epsilon^{\bar a b} & 0
  \end{pmatrix} \quad \text{with} \quad
  \epsilon^{a \bar b} = (\epsilon^T)^{\bar b a}\,.
\end{equation}
Therefore, the small fluctuations initially introduced in the string field $\Psi$ in \eqref{eqn:stringfield} can be thought of parameterizing $\epsilon^{AB}$. These are $D^2$ different entries and allows us to express 
$\mathcal{H}^{AB}$ in a series expansion
\begin{equation}\label{eqn:genmetricexp}
  \mathcal{H}^{AB} = S^{AB} + \epsilon^{AB} + \frac{1}{2} \epsilon^{AC}\, S_{CD}\, \epsilon^{DB} + \dots = \exp ( \epsilon^{AB} )\,.
\end{equation}
Guided by the flux formulation of toroidal DFT \cite{Geissbuhler:2013uka,Aldazabal:2013sca}, let us define the generalized Lie derivative of DFT$_{\rm WZW}$ as
\begin{empheq}[box=\fbox]{equation}
  \mathcal{L}_\lambda V^{A} =  \lambda^B D_B V^{A} + \big( D^A \lambda_B -
  D_B \lambda^A \big)\, V^{B} + F^A{}_{BC} \lambda^B V^C \label{eqn:genLieVA}\,.
\end{empheq}
Objects transforming like $\delta_\lambda V^A= \mathcal{L}_\lambda V^{A}$ are called generalized vectors. The generalized Lie derivative extends to tensors in the usual way so that e.g. the generalized Lie derivative of  $\epsilon^{AB}$ reads
\begin{align}
  \mathcal{L}_\lambda \epsilon^{AB} &=  \lambda^C D_C \epsilon^{AB} + ( D^A \lambda_C - D_C \lambda^A ) \epsilon^{CB} + \nonumber \\ 
  & ( D^B \lambda_C - D_C \lambda^B ) \epsilon^{AC} + F^A{}_{CD} \lambda^C \epsilon^{DB} + F^B{}_{CD} \lambda^C \epsilon^{AD} \label{eqn:genLieepsilonAB}\,.
\end{align}
Moreover, it leaves $\eta^{AB}$ invariant
\begin{equation}
  \mathcal{L}_\lambda \eta^{AB} = 0
\end{equation}
and for a closed gauge parameter it acts trivially, i.e. 
\begin{equation}
\label{liederivtrivial}
  \mathcal{L}_{D^A \chi} V^{B} = 0
\end{equation}
after applying the strong constraint \eqref{eqn:strongconst}. The gauge transformations \eqref{eqn:gaugetrafo} affect fluctuations only. They are trivial 
\begin{equation}\label{eqn:trafobackground} 
  \delta_\lambda S^{AB} = 0 \,
\end{equation}
for the background metric. A straightforward computation shows that the gauge transformation of $\epsilon^{AB}$ can be expressed in terms of the generalized Lie derivative as
\begin{equation}\label{eqn:deltaepsilongenLie}
  \delta_\lambda \epsilon^{AB} = \frac{1}{2} \bigl( \mathcal{L}_\lambda S^{AB} + \mathcal{L}_\lambda \epsilon^{AB} + \mathcal{L}_\lambda S^{(A}{}_C S^{B)}{}_D\,  \epsilon^{CD} \bigr)\,.
\end{equation}
With \eqref{eqn:trafobackground}, one can evaluate the gauge transformation
of the generalized metric
\begin{align}
  \delta_\lambda \mathcal{H}^{AB} &= \delta_\lambda \epsilon^{AB} + \frac{1}{2} \delta_\lambda e^{AC} S_{CD} e^{DB} + \frac{1}{2} e^{AC} S_{CD} \delta_\lambda e^{DB} + \mathcal{O}(\epsilon^2) \\
  &= \frac{1}{2} \bigl( \mathcal{L}_\lambda S^{AB} + \mathcal{L}_\lambda \epsilon^{AB} + \mathcal{L}_\lambda S^{(A}{}_C S^{B)}_D \epsilon^{CD} + \epsilon^{C(A} S_{CD} \mathcal{L}_\lambda S^{B)D} \bigr) + \mathcal{O}(\epsilon^2) \\
  &= \frac{1}{2} ( \mathcal{L}_\lambda S^{AB} + \mathcal{L}_\lambda \epsilon^{AB} ) + \mathcal{O}(\epsilon^2)
    = \frac{1}{2} \mathcal{L}_\lambda \mathcal{H}^{AB} + \mathcal{O}(\epsilon^2)\,.
\end{align}
Being equivalent to \eqref{eqn:O(D,D)gens}, we applied the identity
\begin{equation}
  S^A{}_C \, \epsilon^{CB} = - S^B{}_C \,\epsilon^{CA}
\end{equation}
in the step from the second to the third line. In a similar vein, the gauge transformation of the generalized dilaton $d$
\begin{equation}
  \delta_\lambda d = \frac{1}{2} \mathcal{L}_\lambda d \quad \text{with} \quad
  \mathcal{L}_\lambda d = \lambda^A D_A d - \frac{1}{2} D_A \lambda^A
\end{equation}
can be expressed by using the generalized Lie derivative for a density. In summary, we obtain the very compact notation for the gauge transformations
\begin{empheq}[box=\fbox]{equation}
  \delta_\lambda \mathcal{H}^{AB} = \frac{1}{2} \mathcal{L}_\lambda \mathcal{H}^{AB}
    \quad \text{and} \quad
  \delta_\lambda d = \frac{1}{2} \mathcal{L}_\lambda d \,.
\end{empheq}

\subsection{The C-bracket}
Let us analyze whether the gauge transformations \eqref{eqn:gaugetrafo}  close to give the algebra of the theory. In CSFT, at cubic order the commutator of two gauge transformations $\delta_{\Lambda_1}$ and $\delta_{\Lambda_2}$ gives another one parameterized by
\begin{equation}
  \Lambda_{12} = [\Lambda_2, \Lambda_1]_0 \, .
\end{equation}
Using the techniques presented in section~\ref{sec:cubicorder}, it is straightforward to evaluate this expression and obtain
\begin{align}
  \lambda_{12\,a} &= -\frac{1}{2} \lambda_1^b\, D_b \lambda_{2,a}
    + \frac{1}{4} \Bigl( \lambda_{1,b} \, D_a \lambda_2^b + \lambda_{1, a}\, D_b \lambda_2^b
    - \lambda_1^{\bar b}\, D_{\bar b} \lambda_{2,a} + \lambda_{2,a} \,\mu_1 + 
    f_{abc} \,\lambda_1^b\, \lambda_2^c \Bigr) \nonumber \\
    & \quad - \frac{1}{8} \lambda_{2, a}\, D_{\bar b} \lambda_1^{\bar b} \, -\, (1 \leftrightarrow 2)\,.
\end{align}
Due to the $\mathds{Z}_2$ symmetry \eqref{eqn:Z2sym}, the equation for the $\lambda_{12\,\bar a}$ has exactly the same form. Note that these commutators hold before the field redefinition of the gauge parameter \eqref{eqn:fieldredefgaugepara} is applied. As explained in section~3.1 of \cite{Hohm:2014xsa}, after the field redefinition, we have to adapt $\lambda_{12\,a}$ according to
\begin{equation}
  \lambda'_{12,a} = \lambda_{12,a} + \biggl( 
    \frac{1}{4}\Big( D_{\bar b} \lambda_{1,a}\; \lambda_2^{\bar b}
      + D_a \lambda_{1,\bar b}\, \lambda_2^{\bar b} \Big)
    + \frac{3}{16}\Big( D_{\bar b} \lambda_1^{\bar b}\; \lambda_{2,a} + D_b \lambda_1^b\; \lambda_{2,a} \Big)
    - (1 \leftrightarrow 2) \biggr)\, .
\end{equation}
In addition, we have to set
\begin{equation}
  \mu = \frac{1}{4} D_a \lambda^a - \frac{1}{4} D_{\bar a} \lambda^{\bar a}\,
\end{equation}
which takes into account the $\mu$ gauge fixing performed in the last
subsection. After removing all terms which are not linear in $\lambda_1$ or $\lambda_2$ (or in both), we obtain the result
\begin{equation}
  \lambda'_{12,a} = -\frac{1}{2} \Big(\lambda_1^b D_b + \lambda_1^{\bar b}\, D_{\bar b}\Big) \lambda_{2,a}
    + \frac{1}{4} \Big( \lambda_{1,b}\,  D_a \lambda_2^b - \lambda_{1,\bar
      b}\,  D_a \lambda_2^{\bar b} 
    - f_{abc}\, \lambda_1^b\, \lambda_2^c\Big) - (1 \leftrightarrow 2)\, .
\end{equation}
For the bared parameter we obtain by the same procedure
\begin{equation}
  \lambda'_{12,\bar a} = -\frac{1}{2} \Big(\lambda_1^b D_b +
  \lambda_1^{\bar b}\,  D_{\bar b} \Big) \lambda_{2,\bar a}
    - \frac{1}{4} \Big( \lambda_{1,b}\, D_{\bar a} \lambda_2^b - \lambda_{1,\bar b}\, D_{\bar a} \lambda_2^{\bar b} 
    - f_{\bar a\bar b\bar c}\, \lambda_1^{\bar b}\, \lambda_2^{\bar c}\Big) - (1 \leftrightarrow 2)\,.
\end{equation}
At linear order, $\lambda$ is equivalent to $\lambda'$ and therefore $\lambda$ can be substituted by $\lambda'$ on the right hand side of these two equations. Using the conventions \eqref{conventa}, \eqref{conventb} and \eqref{conventc}, one can write this result in terms of the double index notation, where it takes the very compact form
\begin{equation}
  \lambda_{12}^A = -\frac{1}{2} \lambda_1^B\, D_B \lambda_2^A
    + \frac{1}{4} \lambda_1^B\, D^A \lambda_{2,B} 
    - \frac{1}{4} F^A{}_{BC} \,\lambda_1^B\, \lambda_2^C - (1\leftrightarrow 2)\,.
\end{equation}
This motivates to introduce the C-bracket of DFT$_{\rm WZW}$ as 
\begin{equation}\label{eqn:cbracket}
  [\lambda_1, \lambda_2]^A_\mathrm{C} := -2\lambda_{12}^A=
\lambda_1^B D_B \lambda_2^A -
    \frac{1}{2} \lambda_1^B D^A \lambda_{2\,B} + \frac{1}{2} F^A{}_{BC} \lambda_1^B \lambda_2^C - (1\leftrightarrow 2)
\end{equation}
which differs essentially in the third term from the expression for DFT presented in \cite{Hull:2009zb}. Furthermore, please keep in mind that the
derivatives appearing in \eqref{eqn:cbracket} do not commute.

At this point we observe that  the C-bracket of DFT$_{\rm WZW}$ can also be expressed in terms of the generalized covariant derivative
\begin{equation}\label{eqn:flatcovderv}
  \nabla_A V^B = D_A V^B + \frac{1}{3} F^B{}_{AC} V^C \quad \text{and} \quad
  \nabla_A V_B = D_A V_B + \frac{1}{3} F_{BA}{}^C V_C \,
\end{equation}
as
\begin{empheq}[box=\fbox]{equation}\label{eqn:cbracketcov}
  [\lambda_1, \lambda_2]_C^A = \lambda_1^B \,\nabla_B \lambda_2^A - \frac{1}{2} \lambda_1^B\, \nabla^A\, \lambda_{2,B} - (1 \leftrightarrow 2) \,.
\end{empheq}
In section~\ref{sec:covderiv}, we will discuss this generalized covariant derivative in more detail. The generalized Lie derivative \eqref{eqn:genLieVA} can also be expressed in terms of the covariant derivative as
\begin{empheq}[box=\fbox]{equation}
  \mathcal{L}_{\lambda} V^A = \lambda^C \nabla_C V^A + (\nabla^A \lambda_C - \nabla_C \lambda^A) V^C\,.
\end{empheq}
Due to the total antisymmetry of the structure coefficients $F_{ABC}$, the weak constraint \eqref{eqn:weakconstflat} when acting on a generalized scalar can also written with the covariant derivative
\begin{equation}\label{eqn:strongconstcov}
  \nabla_A D^A f = \bigl(D_A D^A + \frac{1}{3} F^A{}_{AB} D^B \bigl) f = D_A D^Af \,.
\end{equation}
In the context of the weak/strong constraint, the quantities $\epsilon^{AB}$ and $\lambda^A$ appearing in the string fields are treated as generalized scalars. Thus, e.g. $\nabla_A D^A \lambda^B$ gives rise to
\begin{equation}
  \nabla_A D^A \lambda^B = D_A D^A \lambda^B \quad \text{instead of} \quad
  D_A D^A \lambda^B + \frac{1}{3} F^B{}_{AC} D^A \lambda^C\,.
\end{equation}

\subsection{Closure of gauge algebra}
In this section  we check the closure of the gauge algebra. There are two different ways to prove closure which are completely equivalent. First, one can compute the Jacobiator
\begin{equation}
  J(\lambda_1, \lambda_2, \lambda_2) = [\lambda_1, [\lambda_2, \lambda_3]_\mathrm{C}]_\mathrm{C} +
    [\lambda_3, [\lambda_1, \lambda_2]_\mathrm{C}]_\mathrm{C} +
    [\lambda_2, [\lambda_3, \lambda_1]_\mathrm{C}]_\mathrm{C}
\end{equation}
and impose that it vanishes up to terms parameterizing a trivial gauge transformations. According to \eqref{liederivtrivial}, then the constraint
\begin{equation}
  \mathcal{L}_{J(\lambda_1, \lambda_2, \lambda_3)} V^A = 0
\end{equation}
has to hold. Alternatively, one can show that the commutator of two generalized Lie derivatives closes in the sense that
\begin{equation}\label{eqn:closurecond}
  \mathcal{L}_{[\lambda_1, \lambda_2]_\mathrm{C}} V^A = (\mathcal{L}_{\lambda_1} \mathcal{L}_{\lambda_2} 
    - \mathcal{L}_{\lambda_2} \mathcal{L}_{\lambda_1}) V^A\, .
\end{equation}
Here, we will show this second property of the generalized C-bracket.

In the course of the computation, we make extensive use of the commutator of two covariant derivatives
\begin{equation}
  [\nabla_A,\nabla_B] V_C = R_{ABC}{}^D V_D - T^D{}_{AB} \nabla_D V_C \,
\end{equation}
containing the torsion
\begin{equation}
  T^A{}_{BC} = -\frac{1}{3} F^A{}_{BC}
\end{equation}
and the Riemann curvature
\begin{equation}
  R_{ABC}{}^D = \frac{2}{9} F_{AB}{}^E F_{EC}{}^D\,.
\end{equation}
In calculating the Riemann curvature, we used the Jacobi identity
\begin{equation}\label{eqn:jacobiid}
  F_{AB}{}^E F_{EC}{}^D + F_{CA}{}^E F_{EB}{}^D +
    F_{BC}{}^E F_{EA}{}^D = 0
\end{equation}
for the structure coefficients $F_{AB}{}^C$. Note that both the curvature and the torsion of this generalized covariant derivative do not vanish. Thus, the algebra we consider here can be considered as a generalization of
the one proposed by \cite{Cederwall:2014kxa}\footnote{We thank
  David Berman for bringing this  paper to our
  attention.},  which assumed vanishing torsion. We think that it is remarkable that Cederwall proposed a similar algebra by just considering possible generalizations/extensions of the DFT algebra.

Evaluating the condition \eqref{eqn:closurecond}, one eventually arrives at the expression 
\begin{align}
  \mathcal{L}_{[\lambda_1,\lambda_2]_\mathrm{C}} V^A = (\mathcal{L}_{\lambda_1} & \mathcal{L}_{\lambda_2} 
    - \mathcal{L}_{\lambda_2} \mathcal{L}_{\lambda_1}) V^A \nonumber\\
    &- \frac{1}{3} \bigl( F_{BC}{}^F F_{FD}{}^A + F_{DB}{}^F F_{FC}{}^A + F_{CD}{}^F F_{FB}{}^A\bigr) \,,
\end{align}
where the second line vanishes due to the Jacobi identity \eqref{eqn:jacobiid}. Let us emphasize that this closure result goes beyond what one would expect from the CSFT construction. A priori CSFT at cubic order only forces the $V^A$ independent part of \eqref{eqn:closurecond} to hold \cite{Hohm:2014xsa}. For all terms depending on $V^A$, there are in general corrections and closure is only guaranteed on-shell. However, here we do not face any of these problems. Moreover, for the closure of the usual DFT algebra, the strong constraint was essential for the fluctuations and the background, whereas here one only needs the Jacobi-identity for the background flux.

\subsection{Properties of the generalized covariant derivative}\label{sec:covderiv}
Until now, we did not show that $\nabla_A$ really deserves to be called covariant, i.e. that it satisfies the mandatory compatibility conditions \cite{Hohm:2010xe,Hohm:2011si,Aldazabal:2013sca}:
\begin{itemize}
  \item Compatibility with the frame requires
    \begin{equation}\label{eqn:compframe}
      \nabla_A E_B{}^I = 0\,.
    \end{equation}
    Here the covariant derivative acts on a tensor with both, flat and curved indices. Thus, we have to extend its definition
    \begin{equation}
      \nabla_A E_B{}^I = D_A E_B{}^I + \frac{1}{3} F_{BA}{}^C E_C{}^I + E_A{}^K \Gamma_{KJ}{}^I E_B{}^J = 0
    \end{equation}
    by the curved connection $\Gamma_{IK}{}^J$. We already made acquaintance with it in section~\ref{sec:stringfields} while expressing the weak constraint \eqref{eqn:weakconstcov} in terms of a covariant derivative. Due to \eqref{eqn:compframe}, it is completely determined
    \begin{equation}\label{eqn:GammaIJK}
      \Gamma_{IJ}{}^K = - E^A{}_I E^B{}_J E_C{}^K\frac{1}{3} \bigl( 2 \Omega_{AB}{}^C + \Omega_{BA}{}^C \bigr)
      = -\frac{1}{3} \bigl(2 \Omega_{IJ}{}^K + \Omega_{JI}{}^K \bigr)
    \end{equation}
    in terms of the coefficients of anholonomy $\Omega_{ABC} = D_A E_B{}^I E_{CI}$ and the vielbein $E_A{}^I$.
  \item Compatibility with the invariant metric
    \begin{equation}\label{eqn:compinvmetric}
      \nabla_A \eta_{BC} = D_A \eta_{BC} + F_{BA}{}^D \eta_{DC} + F_{CA}{}^D \eta_{BD} = F_{BAC} + F_{CAB} = 0
    \end{equation}
    is fulfilled due to the total antisymmetry of $F_{ABC}$, a direct consequence of the total antisymmetry of its components $f_{abc}$ and $f_{\bar a\bar b\bar c}$. Split into bared and unbared indices, the non-trivial contributions of \eqref{eqn:compinvmetric} read
    \begin{equation}\label{eqn:compinvmetriccomp}
      f_{bac} + f_{cab} = 0 \quad \text{and} \quad - f_{\bar b \bar a \bar c} - f_{\bar  c\bar a \bar b} = 0\,.
    \end{equation}
  \item Compatibility with the background metric
    \begin{equation}
      \nabla_A S_{BC} = D_A S_{BC} + F_{BA}{}^D S_{DC} + F_{CA}{}^D S_{BD} = 
        F_{BA}{}^D S_{DC} + F_{CA}{}^D S_{BD} = 0
    \end{equation}
    is checked along the same lines as for  $\eta$. The only difference is a plus sign instead of a minus sign in the bared part of \eqref{eqn:compinvmetriccomp}.
  \item Compatibility with integration by parts
    \begin{equation}
      \int d^{2 D} X \,e^{-2 d}\, U \,\nabla_M V^M =
        -\int d^{2 D} X \,e^{-2 d}\, \nabla_M U\; V^M
    \end{equation}
    fixes the trace
    \begin{equation}
      \Gamma_{JI}{}^J = \Gamma_I = -2 \partial_I d
    \end{equation}
    of the curved connection. Employing the relation between curved and flat connections \eqref{eqn:GammaIJK}, unimodularity
    \begin{equation}
      F_{AB}{}^B = \Omega_{AB}{}^B - \Omega_{BA}{}^B = 0 \qquad \Leftrightarrow \qquad
        \Omega_{AB}{}^B = \Omega_{BA}{}^B
    \end{equation}
    and \eqref{eqn:Omegaabb}, linking $\Omega_{AB}{}^{B}$ with the flat derivative of $d$, we obtain
    \begin{equation}
      \Gamma_I = - E^A{}_I \Omega_{AB}{}^B = - 2 E^A_I D_A d = -2 \partial_I d\,.
    \end{equation}
    This proves compatibility with integration by parts.
  \item Let us now consider the generalized torsion of $\nabla_A$. Like for DFT, it is defined as the difference between the usual C-bracket and the C-bracket where the partial derivatives are substituted by the covariant ones. In our case, this leads to
\begin{equation}
  [\lambda_1, \lambda_2]_\mathrm{C}^I - \mathcal{T}^I{}_{JK} \lambda_1^J \lambda_2^K = \lambda_1^J \partial_J \lambda_2^I - \frac{1}{2}\lambda_1^J \partial^I \lambda_{2 J} - (1 \leftrightarrow 2)
\end{equation}
with $[\lambda^B_1, \lambda^C_2]_\mathrm{C}^I=[E^B{}_J \lambda^J_1,E^C{}_K\lambda^K_2]_C^A\, E_A{}^I$. Evaluating this expression by using the compatibility with the frame, results in the non-vanishing torsion
\begin{equation}\label{eqn:gentorsion}
  \mathcal{T}^I{}_{JK} = 2  \Gamma_{[JK]}{}^I + \Gamma^I{}_{[JK]}  = -\frac{1}{3} \Bigl( 2 \Omega_{[JK]}{}^I + 2\Omega^I{}_{[JK]} + \Omega_{[J}{}^I{}_{K]} \Bigr)\,.
\end{equation}
Thus, in contrast to the covariant derivative of toroidal DFT, the generalized torsion of the covariant derivative of DFT$_{\rm WZW}$ does not vanish.
\end{itemize}

\section{About the relation of DFT$_{\rm WZW}$ and DFT}
\label{sec:thoughts}
Closely following the original derivation of DFT from CSFT on a toroidal background, we have derived a third order action and the gauge transformations for a DFT describing fluctuations around the WZW model. Note that, since we are working at string tree level and in a large level limit, the left and right moving sector of the background completely decouples so that at this stage we can straightforwardly extend the formalism to left-right asymmetric backgrounds.

We observed that the usual notions of DFT like a generalized Lie derivative, a C-bracket and the strong constraint receive a natural generalization, which encodes, however, the background fields in an intricate way. Both the frame fields and the fluxes of the background appear in the corresponding relations making the above DFT notions explicitly background dependent.

The original double field theory was claimed to be background independent so that the question arises how DFT$_{\rm WZW}$ and DFT and related. If DFT is indeed background independent, then the schematic relation should hold
\begin{equation}
\label{backgrindep}
S_{\rm DFT}(\overline H+\epsilon)\equiv S_{{\rm DFT}_{\rm WZW}}(\tilde\epsilon)
\end{equation}
i.e. the DFT action expanded around the WZW background $\overline H$ should be physically equivalent to the action of DFT$_{\rm WZW}$. Here we indicated that there might exist a non-trivial map between fluctuation $\epsilon$ in DFT and fluctuations $\tilde\epsilon$ in DFT$_{\rm WZW}$.

In this section, we start to analyze the relation between these two theories. A more exhaustive analysis requires the knowledge of the complete  action of DFT$_{\rm WZW}$ in terms of the finite generalized metric \eqref{eqn:genmetricexp}. The construction of this action is beyond the scope of this paper and is postponed to future research \cite{future}. Therefore, in this section we cannot yet provide a fully conclusive picture but merely collect some indications and observations.

\subsection{Asymmetric WZW models as solutions to DFT}
In this section, we show that the asymmetric WZW models, we used as backgrounds, indeed arise as solutions to traditional DFT in the flux formulation \cite{Geissbuhler:2013uka,Aldazabal:2013sca}. 

First, we face the problem that quantities like the generalized vielbein $E_A{}^I$ or the metric $\eta_{IJ}$ are defined differently in the flux formulation and the theory presented here. Hence, it is not straightforward to compare them. The most obvious difference is that the index structures in both formulations are not the same. In the generalized metric formulation of DFT \cite{Hohm:2010pp} the coordinates and partial derivatives read
\begin{equation}
  X^{\hat M}=(\tilde x_i ,\, \hat x^i)\, ,
     \qquad
  \partial^{\hat M}=(\hat \partial_i ,\, \tilde{\partial}^i)\,. 
\end{equation}
Indices marked with a hat are lowered with the $O(D,D)$ invariant metric
\begin{equation}
    \eta_{\hat M\hat N}=\begin{pmatrix}
      0 & \delta_j^i \\
      \delta^j_i & 0
    \end{pmatrix}
\end{equation}
and for the lower-case ones, like $i,j,k,\dots$, the background metric $g_{ij}$ is used. To relate these quantities to the ones used in DFT$_{\rm WZW}$, we consider the diffeomorphism
\begin{equation}
  \tilde x^i = \frac{1}{\sqrt{2}} ( x^i - x^{\bar i} )
    \qquad \text{and} \qquad
  \hat x^i = \frac{1}{\sqrt{2}} ( x^i + x^{\bar i} )
\end{equation}
which is mediated by the matrices
\begin{equation}\label{eqn:MhatMN}
  M^{\hat M}{}_N = \frac{1}{\sqrt{2}} \begin{pmatrix} g_{ij} & - g_{\bar i \bar j} \\
    \delta_j^i & \delta_{\bar j}^{\bar i} \end{pmatrix} = \frac{\partial X^{\hat M}}{\partial X^N} 
    \quad \text{and} \quad
  M_{\hat M}{}^N = \frac{1}{\sqrt{2}} \begin{pmatrix} g^{ij} & g^{\bar i\bar j} \\
    \delta_i^j & - \delta_{\bar i}^{\bar j} \end{pmatrix} = \frac{\partial X_{\hat M}}{\partial X_N}\,.
\end{equation}
Note that it is not a large gauge transformation of DFT \cite{Hohm:2012gk,Berman:2014jba,Hull:2014mxa}, but an ordinary diffeomorphism in the $2 D$ dimensional doubled space. By construction, it links the invariant metric $\eta^{IJ}$ in DFT$_{\rm WZW}$ with its counterpart in traditional DFT according to
\begin{equation}
  M^{\hat M}{}_I M^{\hat N}{}_J \eta^{IJ} = \begin{pmatrix} g_{ij} - g_{\bar i \bar j} &
    \delta_i^j + \delta_{\bar i}^{\bar j} \\[0.15cm]
    \delta_j^i + \delta_{\bar j}^{\bar i} & g^{ij} - g^{\bar i\bar j}
  \end{pmatrix} = \eta^{\hat M\hat N} \quad \text{if} \quad g_{ij} =
  g_{\bar i\bar j}\, .
\end{equation}
For this relation to hold, it is inevitable that in DFT$_{\rm WZW}$ the metric $g_{\bar i\bar j}$ for the right movers and $g_{ij}$ for the left movers coincide. For geometric backgrounds this condition is fulfilled. As we will explicitly see in this section, tree-level DFT$_{\rm WZW}$ makes sense also for a large class of genuinely non-geometric backgrounds. Thus, from this simple point of view, traditional DFT and DFT$_{\rm WZW}$ can at best only be equivalent for left-right symmetric backgrounds.

Recall that there are only two quantities, which carry  all physically relevant information about the group manifold and can be compared directly. These are the totally antisymmetric generalized fluxes
\begin{align}
  \mathcal{F}_{ABC} = 3 D_{[A} E_B{}^{\hat I} E_{C]\hat I} \\
    \label{eqn:F_A} 
  \mathcal{F}_A = \Omega^B{}_{BA} + 2 D_A d = 0 \,.
\end{align}
Please note that the definition of $\mathcal{F}_{ABC}$ given here is the one used in the flux formulation of DFT. It differs from our definition \eqref{eqn:fabcfromvielbein} by a prefactor so that we have to perform the rescaling
\begin{equation}
  \mathcal{F}_{ABC} = \frac{3}{2} F_{ABC}\, .
\end{equation}
Both, $\mathcal{F}_{ABC}$ and $\mathcal{F}_A$, are constant on a group manifolds. In this respect, they are very similar to generalized Scherk Schwarz compactifications \cite{Aldazabal:2011nj,Grana:2012rr} of traditional DFT\footnote{Similar effects arise in massive type II theories, which were discussed in DFT \cite{Hohm:2011cp}, too}. In this context, they have to fulfill several consistency constraints \cite{Dibitetto:2012rk,Geissbuhler:2013uka}. Besides
\begin{equation}
  F_A = \text{const.}\, , \qquad \text{and} \qquad
  F_{ABC} = \text{const.}\,,
\end{equation}
the most important one is the quadratic constraint
\begin{equation}
  \mathcal{F}_{E[AB} \mathcal{F}^E{}_{C]D} = 0 
\end{equation}
which in our setup corresponds to the Jacobi identity of the Lie algebra $\mathfrak{g}$. Recall that it was mandatory for the closure of the gauge algebra \eqref{eqn:jacobiid} discussed in section~\ref{sec:cbracket}. 

Now, let us check whether the WZW background solves the equation of motion of usual DFT. For left-right symmetric WZW this is of course expected, as the background is a solution already to the supergravity equations of motion. One possible way to derive the DFT equations of motion starts from  the generalized Ricci scalar $\mathcal{R}$, which  in flat indices  reads \cite{Geissbuhler:2013uka}
\begin{equation}
  \mathcal{R} = \mathcal{F}_{ABC} \mathcal{F}_{DEF} \Big( \frac{1}{4}
  S^{AD} \eta^{BE} \eta^{CF} - \frac{1}{12} S^{AD} S^{BE} S^{CF} -
  \frac{1}{6} \eta^{AD} \eta^{BE} \eta^{CF} \Bigr)\, ,
\end{equation}
after taking into account that $F_{ABC}=\text{const.}$ and $F_A$ vanishes. By variation with respect to the flat background metric $S^{AB}$, we obtain the symmetric tensor
\begin{equation}
  \mathcal{K}_{AB} = \frac{1}{4} \mathcal{F}_{ACD} \mathcal{F}_{BEF} \bigl( \eta^{CE} \eta^{DF} - S^{CE} S^{DF} \bigl)
\end{equation}
which, after the projection
\begin{equation}\label{eqn:eomDFT}
  R_{AB} = 2 \bar P_{(A}{}^C \bar P_{B)}{}^D \mathcal{K}_{CD}
  \quad \text{with} \quad
  \bar P_{AB} = \frac{1}{2} (\eta_{AB} + S_{AB}) \quad \text{and} \quad
  P_{AB} = \frac{1}{2} (\eta_{AB} - S_{AB})\,,
\end{equation}
gives rise to the generalized Ricci tensor (see e.g. \cite{Hassler:2014sba} for details). For each solution of the equations of motion, this tensor and the generalized Ricci scalar have to vanish. An alternative way to write the equation of motions make use of an antisymmetric tensor $\mathcal{G}^{[AB]}$ \cite{Geissbuhler:2013uka}. However, we will stick to \eqref{eqn:eomDFT} because it is more convenient for expanding the double indices $A$ and $B$ into their bared and unbared components. For the left-right asymmetric structure coefficients $F_{ABC}$ used in this paper, expanding \eqref{eqn:eomDFT} into components give rise to
\begin{align}
  0 &= \frac{4}{9} F_{ac\bar e} F_{\bar b\bar f d} \eta^{cd} \eta^{\bar e\bar f} \quad \text{and} \\
  0 &= -\frac{16 h^\vee}{27 \alpha' k} D \label{eqn:eomR}\, .
\end{align}
Note that the first equation is automatically satisfied as long as we have a strict separation between left and right movers, i.e.  the structure coefficients
\begin{equation}
  F_{\bar a b c} = F_{\bar a\bar b c} = 0 \quad \text{and all permutations thereof}
\end{equation}
vanish. We note that the second equation \eqref{eqn:eomR} is closely related
to the $k^{-1}$ corrections of the central charge \eqref{eqn:centralcharge}
\begin{equation}
  c = \frac{k D}{k + h^\vee} = D \Bigl( 1 - \frac{h^\vee}{k} \Bigr) +
  \mathcal{O}(k'^{-2})\, .
\end{equation}
In an appropriate number of dimensions, the $k$ independent part is canceled by the ghost contribution, whereas the $k^{-1}$ part is canceled by a linear dilaton. In the $k\to\infty$ limit this correction vanishes. Thus we conclude that \eqref{eqn:eomR} is in perfect agreement with our theory, too. Therefore, at this stage even the left-right asymmetric WZW backgrounds are consistent solutions of usual DFT. Note that such an asymmetric background generically violates the strong constraint of toroidal DFT. To see this, consider the term
\begin{equation}
  \frac{1}{6} \mathcal{F}_{ABC} \mathcal{F}^{ABC} + \mathcal{F}^A \mathcal{F}_A 
\end{equation}
which vanishes under the strong constraint \cite{Geissbuhler:2013uka}. According to \eqref{eqn:F_A}, $\mathcal{F}_A$ is zero and thus we are left with 
\begin{equation}
  F_{ABC} F^{ABC} =  \eta^{ab}  \eta^{cd}  \eta^{ef} F_{ace}
  F_{bdf}-\eta^{\bar a \bar b}  \eta^{\bar  c \bar  d}
  \eta^{\bar  e \bar  cf} F_{\bar  a \bar  c \bar e}
  F_{\bar b \bar d \bar f}
     \ne 0
\end{equation}
for $F_{abc}\ne \pm F_{\bar a \bar b \bar c}$.

Let us close this subsection with a comment related to the background independence of toroidal DFT. For the aforementioned background not satisfying the strong constraint, we cannot find even a local frame so that $\tilde\partial^i .=0$, i.e. the background cannot be described in supergravity. Since the weak constraint of DFT for fluctuations around this background 
\begin{equation}
    \partial_I \partial^I (\overline f +\phi)=\partial_I \partial^I \overline f +\partial_I \partial^I \phi=0
\end{equation}
receives an extra additive contribution $\partial_I \partial^I \overline f\ne 0$, it looks very different from the strong constraint of DFT$_{\rm WZW}$. Therefore, at least from this perspective we do not see any possibility how the background independence relation \eqref{backgrindep} can ever be satisfied. Thus we conjecture that DFT$_{\rm WZW}$ for asymmetric WZW models cannot be described by perturbing toroidal DFT around this background.

However, even for the geometric WZW model, the situation is far from being obvious, as there are some substantial differences between DFT$_{\rm WZW}$ and toroidal DFT. As already mentioned, the metric $\eta^{IJ}$ is constant in DFT, while it is space dependent for DFT$_{\rm WZW}$. Moreover, as opposed to DFT$_{\rm WZW}$ , the generalized covariant derivative of DFT has vanishing torsion. Thus, without a deeper analysis it appears to be   difficult to settle these issues.

\subsection{Uplift of genuinely non-geometric backgrounds}
In the previous subsection we have seen that also asymmetric WZW models are solutions to the equation of motion of toroidal DFT. Moreover, they are very similar to generalized Scherk-Schwarz compactifications of the latter theory. First, they satisfy very similar consistency constraints and second they violate the strong constraint. Therefore, it is natural to suspect that the WZW models provide the fully backreacted solutions corresponding to the minima of the effective scalar potential induced by the Scherk-Schwarz reduction. Note that the latter potential is nothing else than the scalar potential of half-maximally (electrically) gauged supergravity. It is important to keep in mind that here we are only working at string tree-level so that e.g. modular invariance at the one-loop level can easily spoil the existence of such a left-right asymmetric CFT.

Let us elaborate on this for the concrete case of  $d=3$ dimensional internal backgrounds. In this case, the authors of \cite{Dibitetto:2012rk} have classified all consistent backgrounds with constant generalized fluxes explicitly. Considering only the ones which give rise to semisimple gaugings, we are left with the three different possibilities listed in table~\ref{tab:solembedding}. 
\begin{table}[b]
  \centering
  \begin{tabular}{|c|c|c|c|c|}
    \hline
    ID & $M_{nm} \sqrt{k} / \cos \alpha$ & $\tilde M^{nm} \sqrt{k} / \sin \alpha$ & gauging & algebra \\
    \hline\hline
    1  & $\diag(1,1,1,1)$ & $\diag(1,1,1,1)$ & $SO(4)$ & $\mathfrak{su}(2) \times \mathfrak{su}(2)$ \\
    \hline
    2  & $\diag(1,1,1,-1)$ & $\diag(1,1,1,-1)$ & $SO(3,1)$ & $\mathfrak{su}(2) \times \mathfrak{sl}(2)$ \\
    \hline
    3  & $\diag(1,1,-1,-1)$ & $\diag(1,1,-1,-1)$ & $SO(2,2)$ & $\mathfrak{sl}(2) \times \mathfrak{sl}(2)$ \\
    \hline
  \end{tabular}
  \caption{Duality orbits of consistent semisimple gaugings with $d=3$
    internal dimensions and $-\pi/4 < \alpha < \pi/4$. This table is
    an extract from table~6 in \cite{Dibitetto:2012rk} which in addition includes non-semisimple setups.} \label{tab:solembedding}
\end{table}
Each of them describes an orbit of physical inequivalent backgrounds parameterized by a real parameter $\alpha$. It is sufficient to focus on the compact orbit 1 because the other two orbits are only the non-compact generalizations of it. Its structure coefficients read
\begin{equation}\label{eqn:fluxsu(2)embedding}
  F_{abc} = \frac{1}{\sqrt{k}} \sqrt{2} \epsilon_{abc} (\cos\alpha + \sin\alpha) \quad \text{and} \quad
  F_{\bar a\bar b\bar c} = \frac{1}{\sqrt{k}} \sqrt{2} \epsilon_{abc} ( \cos\alpha - \sin\alpha )\,.
\end{equation}
For $\alpha=\pi/2$, they reproduce our prime example, the $S^3$ with $H$-flux and inverse string tension $\alpha'=2$, which is discussed in appendix~\ref{app:su(2)}\footnote{Comparing
  \eqref{eqn:fluxsu(2)embedding} with \eqref{eqn:SU2Fabc}, they differ
  by the imaginary unit $i$. This is due to a different conventions
  used in DFT. Whereas, we have a negative definite $S^{AB}$ with
  signature $(-,\dots,-)$, DFT uses a positive definite one with
  signature $(+,\dots,+)$.}. A T-duality transformation along all internal directions flips the sign of the right movers structure coefficients $F_{\bar a\bar b\bar c}$. It is equivalent to a $-\pi/2$ shift of $\alpha$ and acts as
\begin{equation}
  M_{mn} \leftrightarrow - \tilde M^{mn}
\end{equation}
on the parameters $M_{mn}$ and $\tilde M^{mn}$ of the embedding. Thus, the notion of T-duality presented here completely agrees with the convention in \cite{Dibitetto:2012rk}. Except for $\alpha=0$, all other backgrounds in the orbit do not have a geometric T-dual counterpart. They are called genuinely non-geometric backgrounds and violate the strong constraint of toroidal DFT. To see this, one computes
\begin{equation}
  F_{ABC} F^{ABC} = \frac{24}{k^2} \sin ( 2\alpha ) = 0
    \quad \text{only if} \quad \alpha = \frac{\pi}{2} n
    \quad \text{with} \quad n \in \mathds{Z}\,,
\end{equation}
for orbit one in table~\ref{tab:solembedding}. Only the background T-dual to the $S^3$ with $H$-flux is compatible with the strong constraint. All other backgrounds with $\alpha\ne 0$ in the orbit violate the strong constraint. This finding is reflected by the fluxes
\begin{equation}
  M = \diag( H_{123} , Q_1{}^{23}, Q_2{}^{31}, Q_3{}^{12} ) \quad \text{and} \quad
  \tilde M = \diag( R^{123} , f_{23}{}^1, f_{31}{}^2, f_{12}{}^3 )\,,
\end{equation}
too. For $\alpha\ne0$ we alway find $H$- and $R$-flux at the same time.

Thus we conclude that asymmetric WZW models are candidates for the uplift of genuinely non-geometric backgrounds of toroidal DFT. Until now, this uplift was only studied for locally flat backgrounds in terms of asymmetric orbifolds \cite{Condeescu:2012sp,Condeescu:2013yma}. Here, we found a generalization which also works for curved backgrounds. 

\section{Conclusion and Outlook}\label{sec:conclusion}
In this paper we have investigated the effective theory of a closed string propagation on a group manifold with $H$-flux. We started from a purely geometric setup giving rise to a WZW model with two equivalent Ka\v{c}-Moody algebras for the left and right moving parts of the closed string. For this setup, using CSFT we computed the effective action and its gauge transformations up to cubic order in a large level $k$ limit. Consistency required the introduction of the weak constraint \eqref{eqn:weakconstcurved} implementing the CSFT level-matching condition on the fields. In contrast to toroidal DFT, it contained an additional term which could be written as the connection of a covariant derivative. This covariant derivative also appeared when we calculated a generalized Lie derivative and the corresponding C-bracket. It turned out that this generalized covariant derivative has non-vanishing torsion.

Even without having the complete action in terms of a generalized metric yet, we also started to investigate the relation of the new DFT$_{\rm WZW}$ with traditional DFT. We showed that the coordinates used in both descriptions can be related by an ordinary 2D diffeomorphism, but that the metrics $\eta^{IJ}$ only transform properly for left-right symmetric backgrounds. In this respect, the metric $\eta^{IJ}$ of DFT$_{\rm WZW}$ turned out to be coordinate dependent, indicating a possible connection to the work of Cederwall \cite{Cederwall:2014kxa}. Moreover, we checked that the equations of motion of toroidal DFT were satisfied not only for left-right symmetric (geometric) backgrounds but also for asymmetric ones, where the latter do not satisfy the strong constraint of toroidal DFT. These asymmetric WZW backgrounds only had to fulfill the closure constraint for guaranteeing the closure of the gauge algebra under the new strong constraint
\eqref{eqn:strongconst}.

Despite the fact that supergravity is background independent, even for geometric backgrounds, we could not yet conclusively show that usual DFT expanded around a WZW background is physically equivalent to DFT$_{\rm WZW}$. For non-geometric backgrounds violating the strong constraint of toroidal DFT, we found strong indications that DFT$_{\rm WZW}$ goes beyond toroidal DFT. Finally, we studied a concrete class of such asymmetric backgrounds and conjectured that they are related to minima of Scherk-Schwarz reductions of toroidal DFT. In fact, the asymmetric WZW models provide candidates for their string theory uplift. All these findings suggest that DFT$_{\rm WZW}$ contains structures going beyond toroidal DFT.
In relation to toroidal DFT, we are still at a very early stage of developing the full action of DFT$_{\rm WZW}$. One should learn more about the properties of the generalized metric and then try to find a fully self-consistent action of DFT$_{\rm WZW}$ in terms of the generalized metric. Expanded in fluctuations of the metric, this action should reduce to the third order action derived from CSFT in this paper. We hope to report on this in a future  publication \cite{future}.

Besides these fundamental challenges, DFT in (asymmetric) WZW backgrounds opens up many possibilities to study non-geometric backgrounds. The latter can be found via generalized Scherk-Schwarz reductions of toroidal DFT. However, let us emphasize again that the derivations in this paper are all performed at string tree level so that one should analyze whether the proposed up-lifts of these non-geometric gauged supergravity vacua admit e.g. modular invariant one-loop partition functions. From such an analysis one might also learn something about the construction of non-geometric branes \cite{deBoer:2010ud,deBoer:2012ma,Hassler:2013wsa}. For instance, it is known that the near horizon geometry of $k$ NS5-branes is precisely the $SU(2)$ WZW model plus a linear dilaton. Finally, the implications for  non-commutative and non-associative target space structures, as are expected to arise in non-geometric flux backgrounds \cite{Blumenhagen:2010hj,Lust:2010iy,Blumenhagen:2011ph,Bakas:2013jwa,Blumenhagen:2013zpa}, deserve a renewed study in the framework of DFT$_{\rm WZW}$.

\acknowledgments
We would like to thank David Berman, Pascal du Bosque, Olaf Hohm, Stefano Massai, Flavio Montiel, Warren Siegel and Barton Zwiebach for helpful discussions. This work was partially supported by the ERC Advanced Grant ``Strings and Gravity''(Grant.No. 32004) and by the DFG cluster of excellence ``Origin and Structure of the Universe''.

\appendix
\section{The toy model $SU(2)$}\label{app:su(2)}
A nice toy model is the group manifold $SU(2)$ which corresponds to a $S^3$ with $H$-flux. On this background we compute now all relevant quantities discussed through the paper. We start with the generators
\begin{equation}
  t_a = \frac{1}{\alpha' k} \sigma_a \quad \text{with} \quad a = 1, 2, 3
\end{equation}
in the fundamental representation. Here, $\sigma_a$ denote the Pauli-matrices
\begin{equation}
  \sigma_1 = \begin{pmatrix} 0 & 1 \\ 1 & 0 \end{pmatrix}\,, \quad
  \sigma_2 = \begin{pmatrix} 0 & -i \\ i & 0\end{pmatrix}\,, \quad
  \sigma_3 = \begin{pmatrix} 1 & 0 \\ 0 & -1\end{pmatrix} \quad \text{and} \quad
  \sigma_0 = \begin{pmatrix} 1 & 0 \\ 0 & 1 \end{pmatrix}\,.
\end{equation}
The normalization of the generators is chosen in such a way that, according to \eqref{eqn:defetaab}, they give rise to the Killing metric
\begin{equation}
  \eta_{ab} = - \frac{\alpha' k}{2} \frac{\Tr(t_a t_b)}{2 x_f} = \diag (-1 , -1 , -1)
    \quad \text{with} \quad x_f = \frac{1}{2}
\end{equation}
denoting the Dynkin index of the fundamental representation. Each group element
\begin{equation}
  g =  y^0 \sigma_0 - i y^a \sigma_a
\end{equation}
is parameterized in terms of four coordinates $y^i$ which have to fulfill
\begin{equation}
  (y^1)^2 + (y^2)^2 + (y^3)^2 + (y^4)^2 = 1\,.
\end{equation}
Doing so they describe the embedding of a unit three-sphere $S^3$ into the four dimensional euclidean space $\mathds{R}_4$. To parameterize the sphere, we choose Hopf coordinates $x^i = (\eta^1, \eta^2, \eta^3)$ with
\begin{align}
  y^0 &= \cos \eta^2 \cos \eta^1 &
  y^1 &= \sin \eta^2 \cos \eta^1 \\
  y^2 &= \cos \eta^3 \sin \eta^1 &
  y^3 &= \sin \eta^3 \sin \eta^1 \,.
\end{align}
After this preparation, we apply \eqref{eqn:vielbein} and \eqref{eqn:vielbeinbared} to obtain the vielbeins
\begin{align}
  e^a{}_i &= - i \sqrt{k \alpha'}
    \begin{pmatrix}
      0 & \cos^2 \eta^1 & \sin^2 \eta^1 \\
      \cos\eta^{23}_+ & \sin\eta^1 \cos\eta^1 \sin\eta^{23}_+ & 
      -\sin\eta^1 \cos\eta^1\sin\eta^{23}_+  \\
      \sin\eta^{23}_+ & - \sin\eta^1 \cos\eta^1 \cos\eta^{23}_+ &
      \sin\eta^1 \cos\eta^1 \cos\eta^{23}_+
    \end{pmatrix} \quad \text{and} \\
  e^{\bar a}{}_{\bar i} &= -i \sqrt{k \alpha'}
    \begin{pmatrix}
      0 & \cos^2 \eta^1 & -\sin^2 \eta^1 \\
      \cos\eta^{23}_- & \sin\eta^1 \cos\eta^1 \sin\eta^{23}_- & 
      -\sin\eta^1 \cos\eta^1\sin\eta^{23}_-  \\
      -\sin\eta^{23}_- & \sin\eta^1 \cos\eta^1 \cos\eta^{23}_- &
      \sin\eta^1 \cos\eta^1 \cos\eta^{23}_-
    \end{pmatrix} 
\end{align}
with the abbreviation $\eta^{23}_\pm = \eta^2 \pm \eta^3$. They give rise to the structure coefficients \eqref{eqn:fabcfromvielbein}
\begin{equation}\label{eqn:SU2Fabc}
  F_{abc} = \frac{2 i}{\sqrt{\alpha' k}} \epsilon_{abc}
    \quad \text{and} \quad
  F_{\bar a\bar b\bar c} = -\frac{2 i}{\sqrt{\alpha' k}} \epsilon_{abc}
\end{equation}
which, as expected for a geometric background, fulfill $F_{abc} = - F_{\bar a\bar b\bar c}$. The target space metric obtained form the vielbein $e^a{}_i$ reads
\begin{equation}
  g_{ij} = \alpha' k \diag ( 1, \cos^2 \eta^1, \sin^2 \eta^1 )\,.
\end{equation}
It belongs to a $S^3$ with the radius $R = \sqrt{\alpha' k}$. With the structure coefficients \eqref{eqn:SU2Fabc}, \eqref{eqn:WZWHflux} and \eqref{eqn:WZWHfluxexplicit}, we calculate the 3-form 
\begin{equation}
  H = 2\alpha' k\, \sin \eta^1 \cos \eta^1 d\eta^1 \wedge d\eta^2 \wedge d\eta^3 \,.
\end{equation}
As a consistency check we evaluation the quantization condition
\begin{equation}
  \frac{1}{2\pi \alpha'} \int_{S^3} H = \frac{k}{\pi} \int\limits_0^{2\pi} d\eta^2 
    \int\limits_0^{2\pi} d\eta^3 \int\limits_0^{\pi/2} d\eta^1\,\sin \eta^1 \cos\eta^1 =
      2 \pi k
\end{equation}
for the $H$-flux. It reproduces the quantization condition $k\in\mathds{N}$ for the level on compact group manifolds.

Following the prescription outlined in section~\ref{sec:representation}, one obtains the functions
\begin{center}
  \begin{tabular}{l|cccc}
    $y_{\lambda q}$ & $\lambda = 0$ & $\lambda = 1/\sqrt{2}$ & $\lambda=\sqrt{2}$ & $\cdots$ \\
    \hline\hline
    $\vdots$ & & & &  $\iddots$ \\
    $q=\sqrt{2}$ & & & 
      $\displaystyle \frac{\sqrt{3} e^{i 2 \eta^3} \sin^2 \eta^1}{\sqrt{2} \pi (\alpha'k)^{3/4}}$ & $\cdots$ \\
    $q=\displaystyle \frac{1}{\sqrt{2}}$ & &
      $ \displaystyle \frac{ e^{i \eta^3} \sin \eta^1}{\pi (\alpha'k)^{3/4}}$ &
      -- & $\cdots$ \\
      $q=0$ & 0 & -- & $\displaystyle - \frac{ \sqrt{3} e^{i (\eta^3 - \eta^2)} \cos \eta^1 \sin \eta^1}{\pi (\alpha'k)^{3/4}}$ & $\cdots$ \\
    $q=\displaystyle -\frac{1}{\sqrt{2}}$ & & 
      $-\displaystyle\frac{e^{-i\eta^2} \cos\eta^1}{\pi (\alpha'k)^{3/4}}$ & -- & $\cdots$ \\
    $q=-\sqrt{2}$ & & &
      $\displaystyle \frac{\sqrt{3} e^{-i 2 \eta^2} \cos^2 \eta^1}{\sqrt{2} \pi (\alpha' k)^{3/4}}$ & $\cdots$ \\
      $\vdots$ & & & & $\ddots$
  \end{tabular}
\end{center}
which form an orthonormal basis of the Hilbert space of square-integrable functions on the $S^3$.

\section{Geometry of the three-point string vertex}\label{app:3stringvertex}
The quadratic differential for the three-punctured sphere with punctures at $z_{i\,0} = (\infty,0,1)$ reads \cite{Witten:1985cc,Sonoda:1989wa}
\begin{equation}
  \varphi(z) = \phi(z) (d z)^2 \quad \text{with} \quad 
    \phi(z) = - \frac{1}{(z-1)^2} - \frac{1}{z^2} + \frac{1}{z (z+1)}\,.
\end{equation}
A local coordinates system around these punctures reproducing $\varphi(z)$ is given in terms of the functions $z = f_i(z_i)$ with the property
\begin{equation}\label{eqn:localcoords}
  \frac{d f_i}{d z_i} = \sqrt{\phi(z_i)}\,.
\end{equation}
Expanding the left and right hand side of this equation into a Laurent series around $z_{2\,0}=0$, it is 
straightforward to show that the function
\begin{equation}
  f_2(z_2) = \frac{(\sqrt{3} - i) \bigl[ (i + z)^{2/3} + (i - z)^{2/3} \bigr]}{%
    (\sqrt{3} + i) (i + z)^{2/3} + 2 i(i - z)^{2/3}}
\end{equation}
is a solution of \eqref{eqn:localcoords}. A Taylor expansion of $f_2(z_2)$ around $z_{2,0}$ gives rise to
\begin{equation}
  f_2(z_2) = -\frac{4}{3\sqrt{3}} z_2 - \frac{8}{27} z_2^2 + \frac{4}{81\sqrt{3}} z_2^3 + \frac{16}{243} z_2^4 - \frac{52}{2187\sqrt{3}} z_2^5 + \cdots \,.
\end{equation}
We compare this expansion with \eqref{eqn:f_2expansion} and finally obtain
\begin{equation}
  \rho = -\frac{4}{3\sqrt{3}}\,, \quad
  d_1 = -\frac{1}{2} \,, \quad
  d_2 = -\frac{1}{16} \,, \quad
  d_3 = \frac{3}{16} \,, \quad
  d_4 = \frac{13}{256} \,, \quad \dots
\end{equation}
as stated in \eqref{eqn:coeff_2expansion}. The remaining functions $f_3(z_3)$ and $f_1(z_1)$ arise from the M\"obius transformations
\begin{equation}
  z \rightarrow \frac{1}{1 - z} \quad \text{and} \quad
  z \rightarrow 1 - \frac{1}{z}
\end{equation}
which permute the punctures of the sphere. 

\bibliography{literatur}
\bibliographystyle{JHEP}

\end{document}